\documentclass[12pt]{extarticle}
\usepackage{slashed}
\usepackage{color,colortbl,verbatim}
\usepackage{latexsym}
\usepackage{amssymb}
\usepackage{amsmath}
\usepackage{graphicx}
\graphicspath{{Figures/}}
\usepackage{arydshln}
\usepackage[dvipsnames]{xcolor}
\usepackage{adjustbox}
\usepackage{easybmat}
\usepackage{bbm}
\usepackage{cite}
\usepackage{slashed}
\usepackage{bm}
\usepackage{adjustbox}
\usepackage{booktabs}
\usepackage{multirow} 
\usepackage{rotating}
\usepackage{mathtools}
\usepackage[inline]{enumitem}
\usepackage{lscape}
\usepackage{booktabs}
\usepackage[normalem]{ulem} 
\usepackage{hyperref}
\usepackage{bm}
\newcommand{\ii}{\ensuremath{\mathrm{i}}}

\newcommand{\be}{\begin{equation}}
\newcommand{\ee}{\end{equation}}
\newcommand{\bea}{\begin{eqnarray}}
\newcommand{\eea}{\end{eqnarray}}

\RequirePackage[normalem]{ulem} 
\RequirePackage{color}\definecolor{RED}{rgb}{1,0,0}\definecolor{BLUE}{rgb}{0,0,1} 

\definecolor{LightCyan}{rgb}{0.88,1,1}
\definecolor{LightGray}{rgb}{0.8,0.8,0.8}
\newcommand{\mc}[1]{#1}

\begin{document}
\thispagestyle{empty}
\begin{flushright}
\end{flushright}
\vspace{0.8cm}

\begin{center}
{\Large\sc Positivity restrictions on the mixing of \\[0.3cm] dimension-eight SMEFT operators}
\vspace{0.8cm}

\textbf{
Mikael Chala$^{\,\bm{a}}$ and Xu Li$^{\,\bm{b,c}}$
}\\
\vspace{1.cm}
{$^a\,\,$\em {Departamento de F\'isica Te\'orica y del Cosmos,
Universidad de Granada, Campus de Fuentenueva, E--18071 Granada, Spain}\\[0.5cm]
$^b\,\,$ {Institute of High Energy Physics, Chinese Academy of Sciences, Beijing 100049, China}\\[0.5cm]
$^c\,\,$ {School of Physical Sciences, University of Chinese Academy of Sciences, Beijing 100049, China}
}
\vspace{0.5cm}
\end{center}
\begin{abstract}
We discuss the structure of the mixing among dimension-eight operators in the
SMEFT relying on the positivity of two-to-two forward scattering amplitudes.
We uncover tens of new non-trivial zeros as well as hundreds of terms with definite sign in (a particular basis of)
the corresponding anomalous dimension matrix. We highlight that our results are not
immediately apparent from the Feynman diagrammatic perspective, nor from on-shell
amplitude
methods. As a byproduct of this work, we provide positivity bounds not
previously derived in the literature, as well as explicit values of certain
elements of the anomalous dimension matrix that serve for cross-check of our
results. 
%
%
%
\end{abstract}

\newpage


\section{Introduction}
\label{sec:Introduction}
The SMEFT is the extension of the Standard Model (SM) with operators of dimension higher than four, suppressed by powers of the cutoff $\Lambda$; see Refs.~\cite{Brivio:2017vri,Isidori:2023pyp} for recent reviews. During the last decades, it has increasily become one of the most popular descriptions of physics beyond the SM, with different theoretical aspects of this framework being better and better understood. Among the most important ones, we could highlight the following three.

(i) The number of independent physical operators up to dimension 15 has been computed using different techniques, including Hilbert series methods~\cite{Henning:2015alf,Kondo:2022wcw}, standard group theory  techniques~\cite{Criado:2019ugp,Fonseca:2019yya} or on-shell amplitudes~\cite{AccettulliHuber:2021uoa,DeAngelis:2022qco,Ren:2022tvi}. The explicit form of these operators have been also computed in several instances, and different software tools (partially) automatising this construction are already available~\cite{Criado:2019ugp,Fonseca:2019yya,Li:2022tec,Harlander:2023psl}. Similar considerations have been also investigated for operators independent off-shell~\cite{Gherardi:2020det,Chala:2021cgt,Ren:2022tvi,Zhang:2023kvw}.

(ii) The physical parameter space of the SMEFT, particularly at dimension eight, has been considerably constrained in light of positivity bounds. These are restrictions on the sign of Wilson coefficients ensuing from very robust principles, including the analiticity, unitarity and the large-energy behaviour of the S-matrix~\cite{Adams:2006sv}. Many of these constraints are strongly competitive with experimental limits~\cite{Zhang:2018shp,Bellazzini:2018paj,Bi:2019phv,Remmen:2019cyz,Zhang:2020jyn,Remmen:2020vts,Fuks:2020ujk,Gu:2020ldn,Chala:2021wpj,deRham:2022sdl,Ghosh:2022qqq,Kim:2023pwf,Ellis:2023zim,Altmannshofer:2023bfk,Kim:2023bbs}.

(iii) The full renormalisation group equations (RGEs) of the SMEFT are known up to dimension six~\cite{Antusch:2001ck,Jenkins:2013zja,Jenkins:2013wua,Alonso:2013hga,Adams:2006sv}, including finite terms (or matching corrections) induced by evanescent operators~\cite{Fuentes-Martin:2022vvu}; as well as to dimension seven~\cite{Liao:2016hru,Liao:2019tep,Chala:2021juk,Zhang:2023kvw}. And the matching of different models onto the SMEFT to this order has been also investigated~\cite{Jiang:2018pbd,Gherardi:2020det,Chala:2020vqp,Haisch:2020ahr,Dittmaier:2021fls,Zhang:2021jdf,Fuentes-Martin:2020udw,Carmona:2021xtq,Fuentes-Martin:2022jrf,DeAngelis:2023bmd}.

Despite the significant progress made also towards renormalising the SMEFT to dimension eight~\cite{Chala:2021pll,AccettulliHuber:2021uoa,DasBakshi:2022mwk,Helset:2022pde,Ardu:2022pzk,Asteriadis:2022ras,DasBakshi:2023htx,Assi:2023zid}, this endevour is still far from complete, due mostly to the enourmous amount of operators~\cite{Murphy:2020rsh,Li:2020gnx}. However, there is a reasonable understanding of the structure of mixings based on generalised unitarity and on-shell amplitude techniques, which highlight certain non-renormalisation results otherwise obscure within the realm of Feynman diagrams~\cite{Cheung:2015aba,Shadmi:2018xan,Ma:2019gtx,Bern:2019wie,Craig:2019wmo,Jiang:2020rwz}. More recently, it has been proposed~\cite{Chala:2023jyx} that the sign of some mixing terms, as well as certain zeros that are not even apparent from the on-shell amplitude perspective, can be unveiled upon studying the positivity of two-to-two scattering amplitudes in the forward limit at very large distances. The basic idea is that, as indicated before, certain Wilson coefficients $c_i$ are constrained to be $c_i\geq 0$. For this to hold irrespective of the running induced by any other Wilson coefficient $c_j$ at a scale $\mu$, $c_i(\mu)\sim \gamma_{ij} c_j\log{\mu/\Lambda}$, where $\gamma$ is the so-called anomalous dimension, the condition $\gamma_{ij}\leq 0$ must be satisfied if $c_j$ is also bounded by positivity, $c_j\geq 0$; otherwise $\gamma_{ij}=0$. 

In Ref.~\cite{Chala:2023jyx}, a precise formulation of this idea was applied to the electroweak and leptonic sectors of the SMEFT. In this paper, we extend the methods of that reference to the full SMEFT including quarks and colour. We organise the article as follows. In section~\ref{sec:theory}, we describe the structure of the SMEFT and provide an example of vanishing mixing that is neither apparent from the Feynman-diagrammatic approach nor from the point of view of on-shell amplitude methods. In section~\ref{sec:positivity}, we discuss briefly the constraints on anomalous dimensions implied by the positivity of the S-matrix, making special emphasis on the derivation of complete positivity bounds. We apply these methods to uncover the structure of the anomalous dimension matrix (ADM) of the SMEFT in section~\ref{sec:structure}. Finally, we conclude in section~\ref{sec:conclusions}. Appendices~\ref{app:comparison}, \ref{app:bounds}, \ref{app:adms} and \ref{app:xchecks} complement the previous discussions.

\section{Theoretical background}
\label{sec:theory}
Our convention for the SM Lagrangian is the following:
\begin{align}
 \mathcal{L}_\text{SM} &= -\frac{1}{4}G_{\mu\nu}^A G^{A\,\mu\nu}-\frac{1}{4}W^I_{\mu\nu}W^{I\,\mu\nu}-\frac{1}{4}B_{\mu\nu}B^{\mu\nu}\\\nonumber
 &+\overline{q}\ii\slashed{D}q+\overline{l}\ii\slashed{D} l+\overline{u}\ii\slashed{D}u+\overline{d}\ii\slashed{D}d+\overline{e}\ii\slashed{D}e\\\nonumber
 &+ (D_\mu\phi)^\dagger (D^\mu\phi)+\mu_\phi^2|\phi|^2 -\lambda_\phi |\phi|^4- (\overline{q} \tilde{\phi} Y_u u + \overline{q}\phi Y_d d+\overline{l}\phi Y_e e + \text{h.c.})\,.
\end{align}
As usual, $G$, $W$ and $B$ denote the $SU(3)_c$, $SU(2)_L$ and $U(1)_Y$ gauge fields with gauge couplings $g_1$, $g_2$ and $g_3$, respectively; $q$ and $l$ stand for the left-handed quarks and leptons, respectively, and $u,d$ and $e$ for the right-handed counterparts; $\phi$ represents the Higgs doublet and $Y_{u}$, $Y_d$ and $Y_e$ are the Yukawa couplings.
We work in the limit of one family of fermions.

The SMEFT Lagrangian to dimension eight, ignoring lepton-number violation, reads:
\begin{equation}
 \mathcal{L}_\text{SMEFT} = \mathcal{L}_\text{SM} + \frac{1}{\Lambda^2} \sum_i c_i^{(6)}\mathcal{O}_i^{(6)} + \frac{1}{\Lambda^4}\sum_j c_j^{(8}\mathcal{O}_{j}^{(8)}\,,
\end{equation}
where $\Lambda$ represents the cutoff of the effective-field theory (EFT). The first and second sums above run over bases of dimension-six~\cite{Grzadkowski:2010es} and dimension-eight~\cite{Murphy:2020rsh,Li:2020gnx} operators, respectively. We are only interested on the dimension-eight ones, for which we use the self-explanatory notation of Ref.~\cite{Murphy:2020rsh}. 
The mixing of these operators under renormalisation group running is governed by the corresponding RGEs:
\begin{equation}
 \dot{c}_i^{(8)} \equiv 16\pi^2\mu\frac{d c_i^{(8)}}{d\mu} = \gamma_{ij}c_j^{(8)} + \gamma_{ijk}' c_j^{(6)} c_k^{(6)}\,.
\end{equation}
The anomalous dimension $\gamma$ characterises the mixing between dimension-eight operators due to SM terms, while $\gamma'$ describes the renormalisation of dimension-eight operators by pairs of dimension-six ones. Our goal in this paper is constraining the shape of $\gamma$ from the positivity of the S-matrix~\cite{Chala:2023jyx}. Previous restrictions on $\gamma$, and in particular the occurrence of certain non-obvious zeros, can be found in Refs.~\cite{Elias-Miro:2014eia,Cheung:2015aba,Bern:2019wie,Bern:2020ikv,Cao:2021cdt,Cao:2023adc}. See also Ref.~\cite{Helset:2022pde} for complementary results based on a geommetric description of the SMEFT.

As a simple motivation for this work, let us consider the example of the mixing of $e^2\phi^2 D^3$ into $e^2 B^2 D$. There are two operators in the first class, and only one in the second; see Ref.~\cite{Murphy:2020rsh}. For convenience, we write them here explicitly:
\begin{align}
 \mathcal{O}_{e^2 B^2 D} &= \ii (\overline{e}\gamma^\mu D^\nu e) B_{\mu\rho} B_{\nu}^{\,\rho}+\text{h.c.}\,\\
 \mathcal{O}_{e^2\phi^2 D^3}^{(1)} &= \ii (\overline{e}\gamma^\mu D^\nu e) (D_{(\mu} D_{\nu)}\phi^\dagger\phi) + \text{h.c.}\,, \\
 \mathcal{O}_{e^2\phi^2 D^3}^{(2)} &= \ii (\overline{e}\gamma^\mu D^\nu e)(\phi^\dagger D_{(\mu} D_{\nu)}  \phi)+\text{h.c.}\,.
\end{align}
\begin{figure}[t]
 \begin{center}
 {\includegraphics[width=0.15\columnwidth]{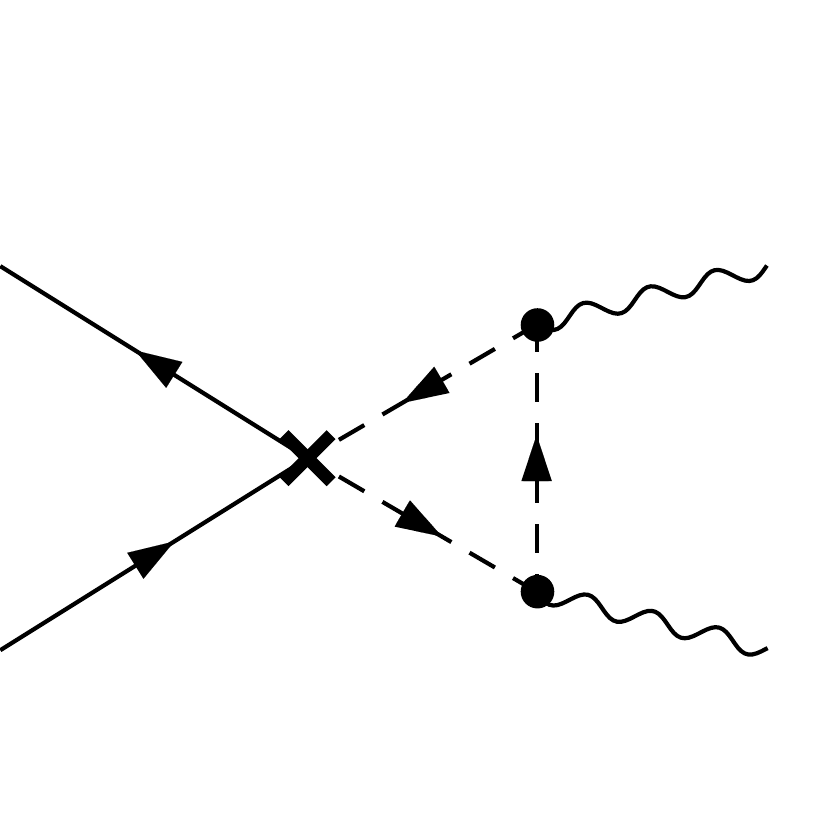}
 \includegraphics[width=0.15\columnwidth]{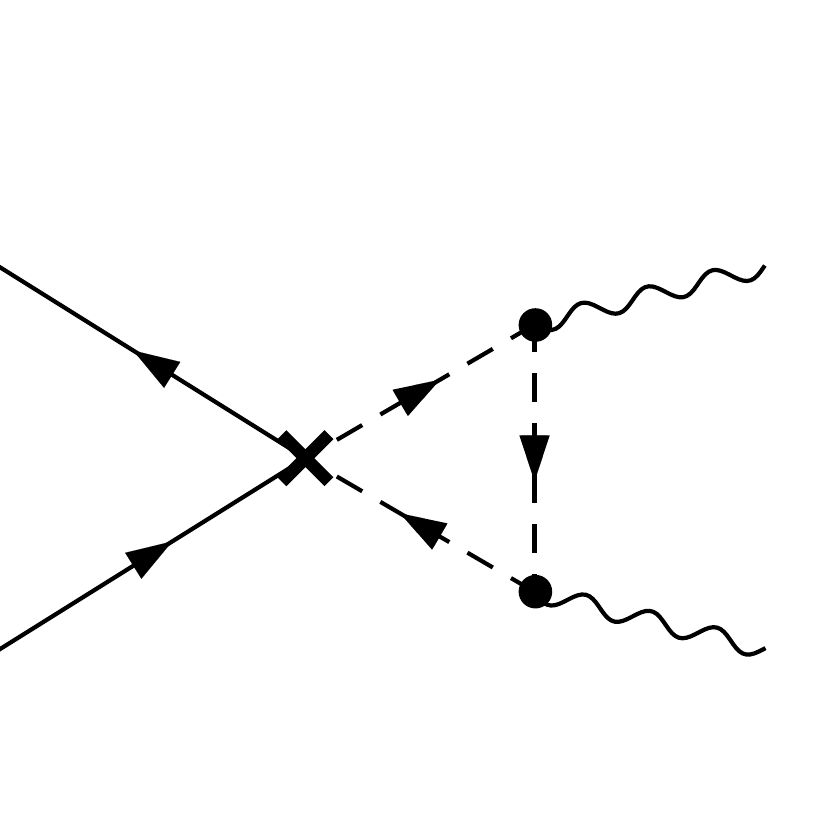}
 \includegraphics[width=0.15\columnwidth]{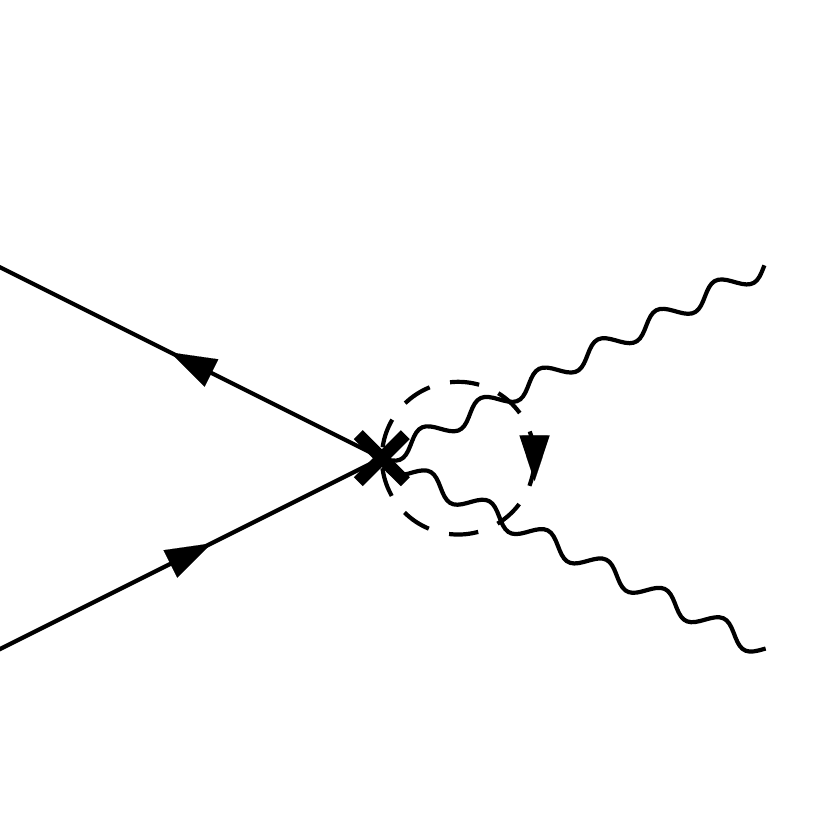}
 \includegraphics[width=0.15\columnwidth]{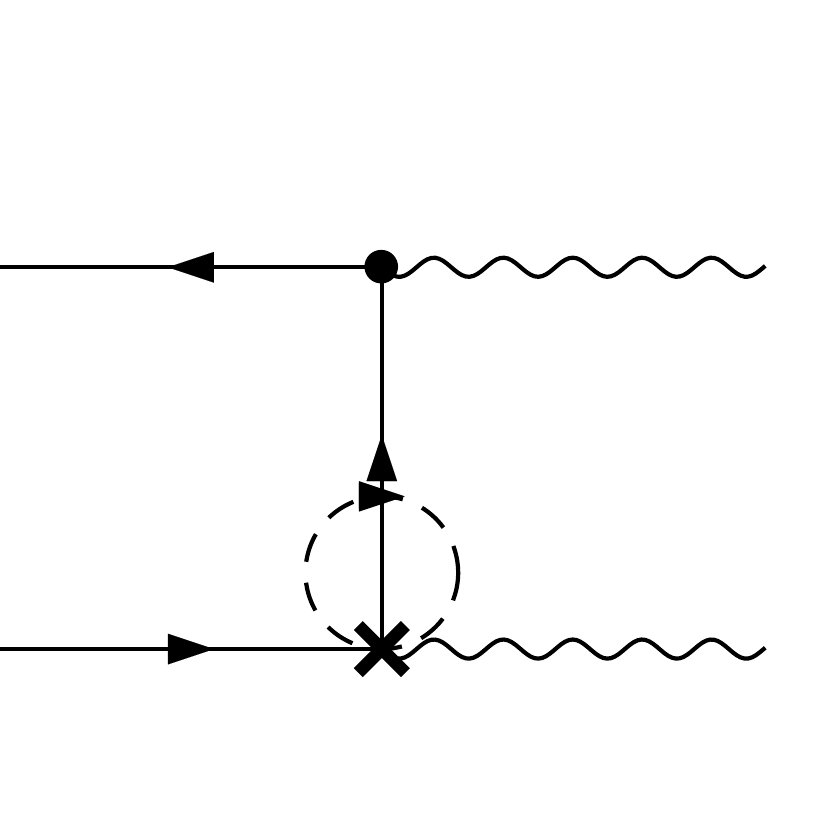}}
 %
 %
 {\includegraphics[width=0.15\columnwidth]{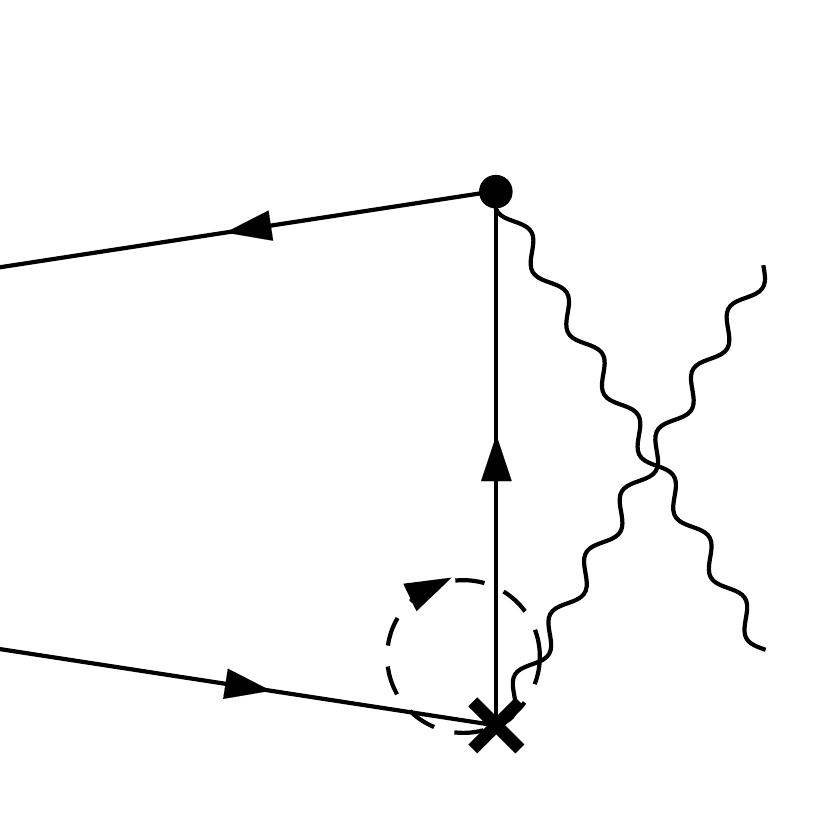}
 \includegraphics[width=0.15\columnwidth]{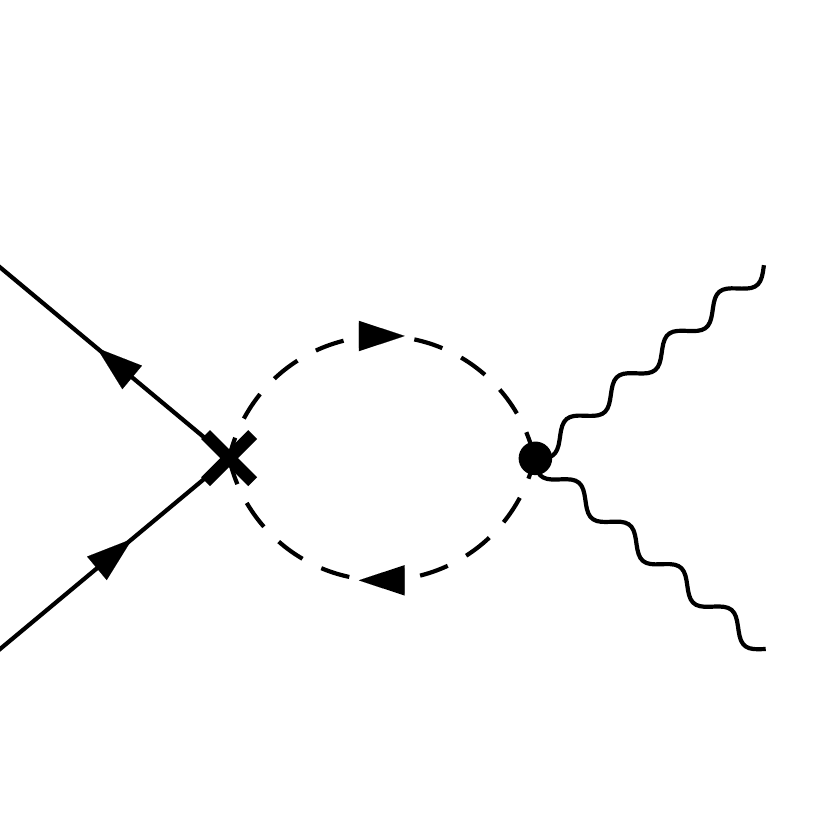}
 \includegraphics[width=0.15\columnwidth]{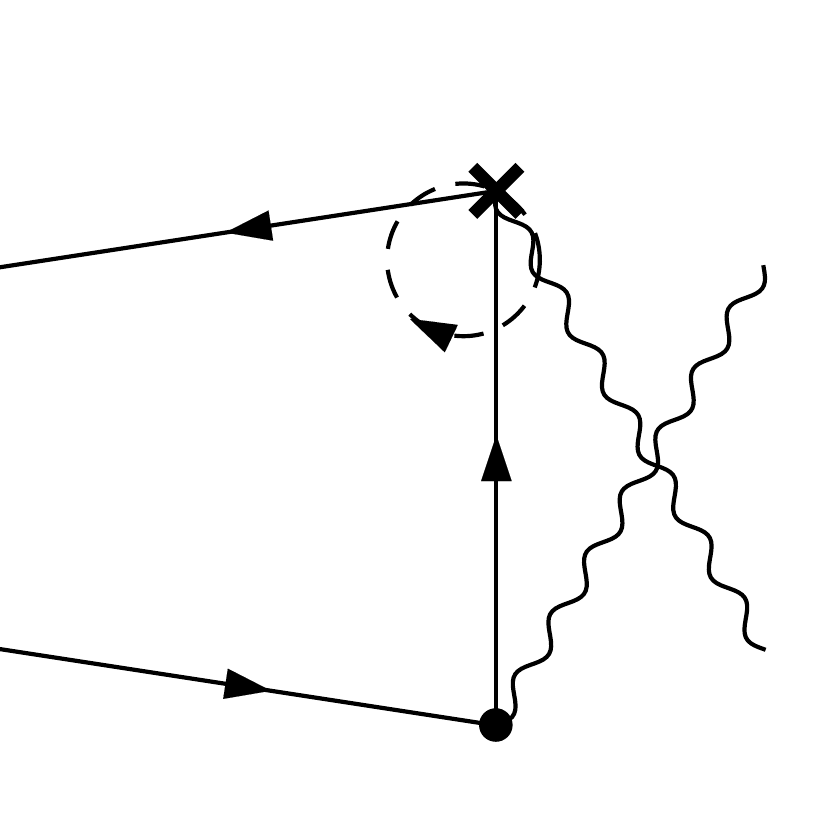}
 \includegraphics[width=0.15\columnwidth]{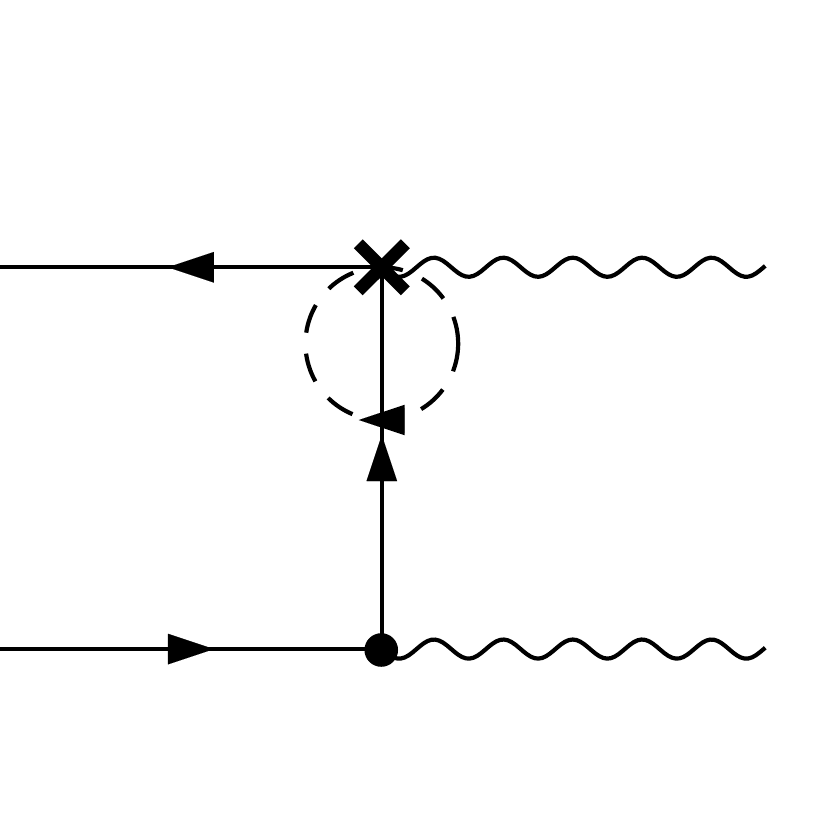}}
 {\includegraphics[width=0.15\columnwidth]{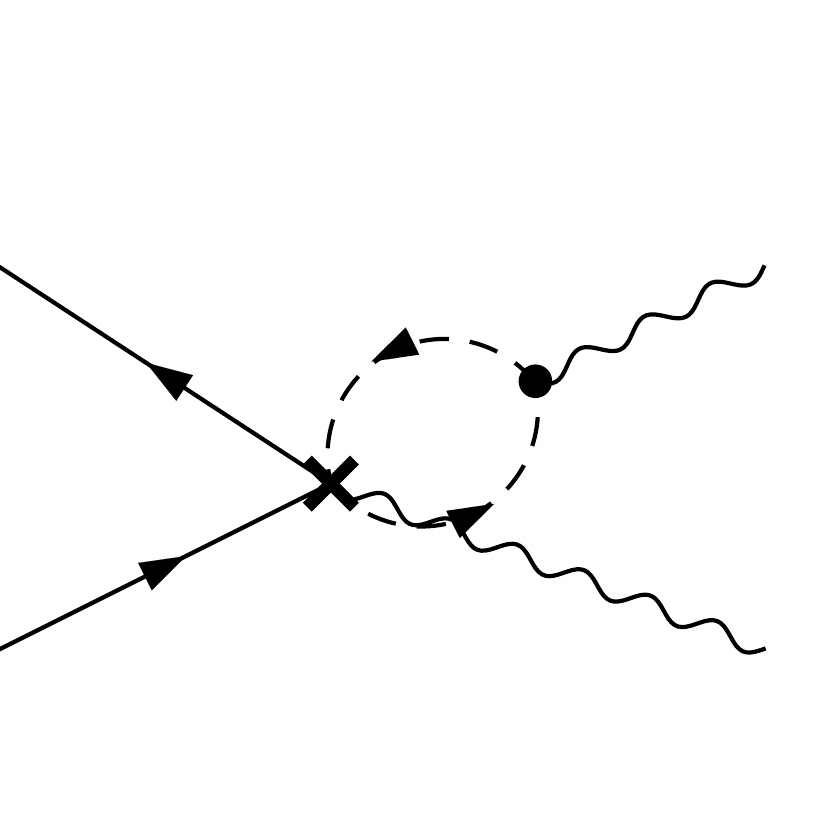}
 \includegraphics[width=0.15\columnwidth]{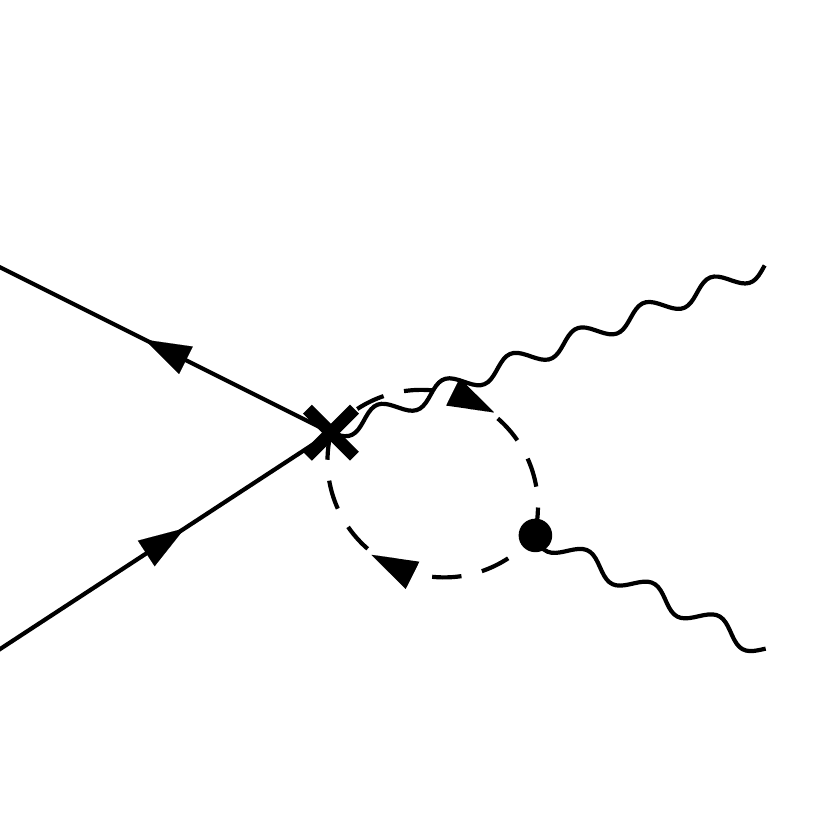}
 \includegraphics[width=0.15\columnwidth]{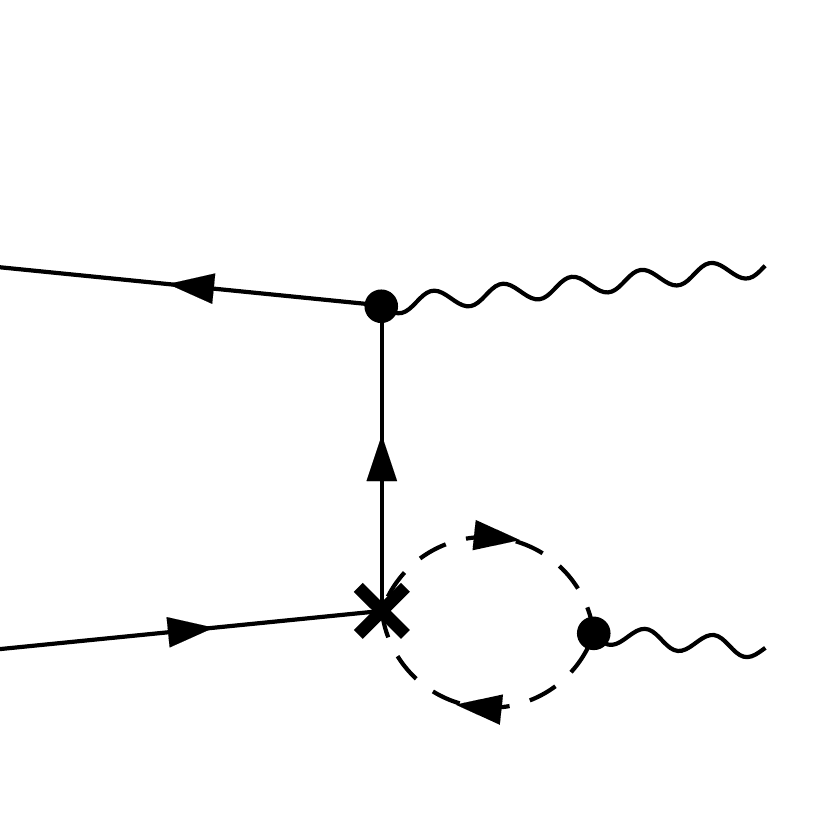}
 \includegraphics[width=0.15\columnwidth]{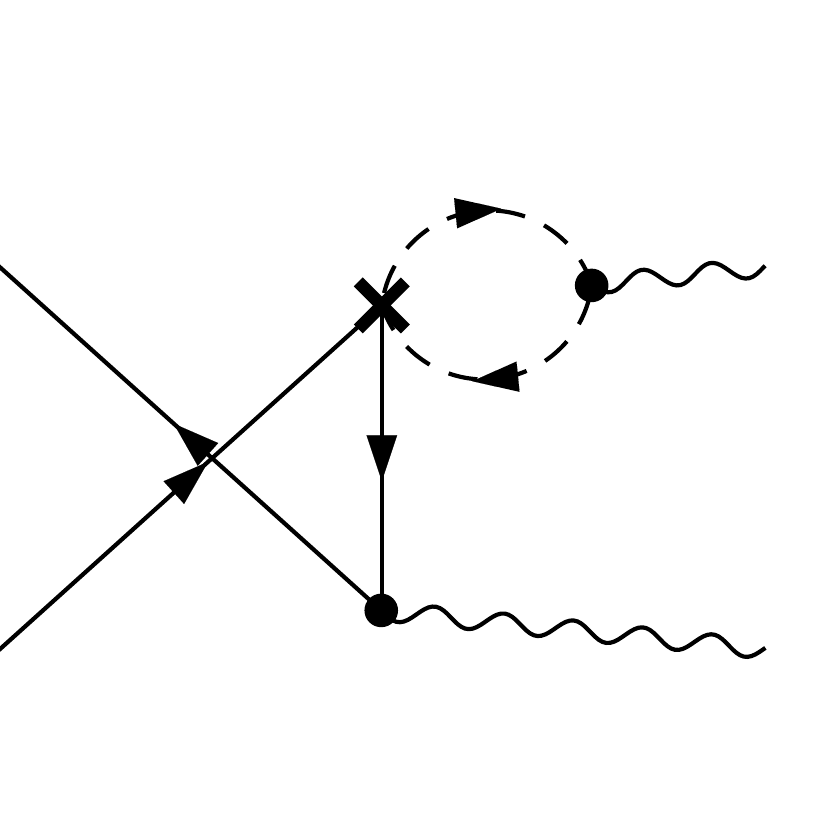}}
 \end{center}
 \caption{\it Feynman diagrams relevant for the one-loop renormalisation of $e^2 B^2 D$ by $e^2\phi^2 D^3$. Diagrams with massless bubbles, which vanish in dimensional-regularisation, are not shown.}\label{fig:diagrams}
\end{figure}
It is obvious that there exists some linear combination of $\mathcal{O}_{e^2\phi^2 D^3}^{(1)}$ and $\mathcal{O}_{e^2\phi^2 D^3}^{(2)}$ that does not mix into $\mathcal{O}_{e^2 B^2 D}$. Indeed, if any of the two anomalous dimensions, say $\gamma_{c_{e^2 B^2 D}, \, c_{e^2\phi^2 D^3}^{(2)}}$, is non-zero, then we can simply define:
\begin{equation}
 \widetilde{O}_{e^2\phi^2 D^3}^{(1)} = \mathcal{O}_{e^2\phi^2 D^3}^{(1)} - \frac{\gamma_{c_{e^2 B^2 D}, \, c_{e^2\phi^2 D^3}^{(1)}}}{\gamma_{c_{e^2 B^2 D}, \, c_{e^2\phi^2 D^3}^{(2)}}} \mathcal{O}_{e^2\phi^2 D^3}^{(2)}\,,
\end{equation}
which by construction does not renormalise $\mathcal{O}_{e^2 B^2 D}$.

The exact form of $\widetilde{\mathcal{O}}_{e^2\phi^2 D^3}^{(1)}$ can be obtained by explicit computation of $\dot{c}_{e^2 B^2 D}$. From the Feynman-diagrammatic perspective, we must compute the diagrams in Fig.~\ref{fig:diagrams}. Using dimensional regularisation with space-time dimension $d=4-2\epsilon$, and with the help of \texttt{FeynArts}~\cite{Hahn:2000kx} and \texttt{FormCalc}~\cite{Hahn:1998yk}, we find that most of them vanish; the rest, despite the many different terms involving a variety of products of momenta, polarizations and gamma matrices, shrink to give the compact result:
\begin{equation}
 \dot{c}_{e^2 B^2 D} = -\frac{1}{3}g_1^2 (c_{e^2\phi^2 D^3}^{(1)}+c_{e^2\phi^2 D^3}^{(2)})\,.
\end{equation}
From here, it is clear that $\widetilde{\mathcal{O}}_{e^2\phi^2 D^3}^{(1)}$ is the linear combination of $\mathcal{O}_{e^2\phi^2 D^3}^{(1)}$ and $\mathcal{O}_{e^2\phi^2 D^3}^{(2)}$ with different signs. In other words, let us make, for example, the basis transformation defined by:
\begin{equation}\label{eq:basistransformation}
 \begin{pmatrix}c_{e^2\phi^2 D^3}^{(1)} \\c_{e^2\phi^2 D^3}^{(2)}\end{pmatrix} = P_{e^2\phi^2 D^3} \begin{pmatrix}\tilde{c}_{e^2\phi^2 D^3}^{(1)}\\\tilde{c}_{e^2\phi^2 D^3}^{(2)}\end{pmatrix}\,,\quad P_{e^2\phi^2 D^3} = \begin{pmatrix}1 & 0\\-1 & 1\end{pmatrix}\,.
\end{equation}
Then, we have that:
\begin{equation}\label{eq:zerofeynman}
 \dot{c}_{e^2 B^2 D}  = -\frac{1}{3}g_1^2 \begin{pmatrix}1 & 1\end{pmatrix}P_{e^2\phi^2 D^3}\begin{pmatrix}\tilde{c}_{e^2\phi^2 D^3}^{(1)}\\\tilde{c}_{e^2\phi^2 D^3}^{(2)}\end{pmatrix} = (\fcolorbox{black}{LightCyan}{$\,0\,$} \,\, -\frac{1}{3}g_1^2)\begin{pmatrix}\tilde{c}_{e^2\phi^2 D^3}^{(1)}\\\tilde{c}_{e^2\phi^2 D^3}^{(2)}\end{pmatrix}\,. 
\end{equation}
What is not obvious at all is how one can anticipate a matrix like $P_{e^2\phi^2 D^3}$ without computing explicitly the anomalous dimensions involved.

\begin{figure}
 \begin{center}
  \includegraphics[width=0.5\columnwidth]{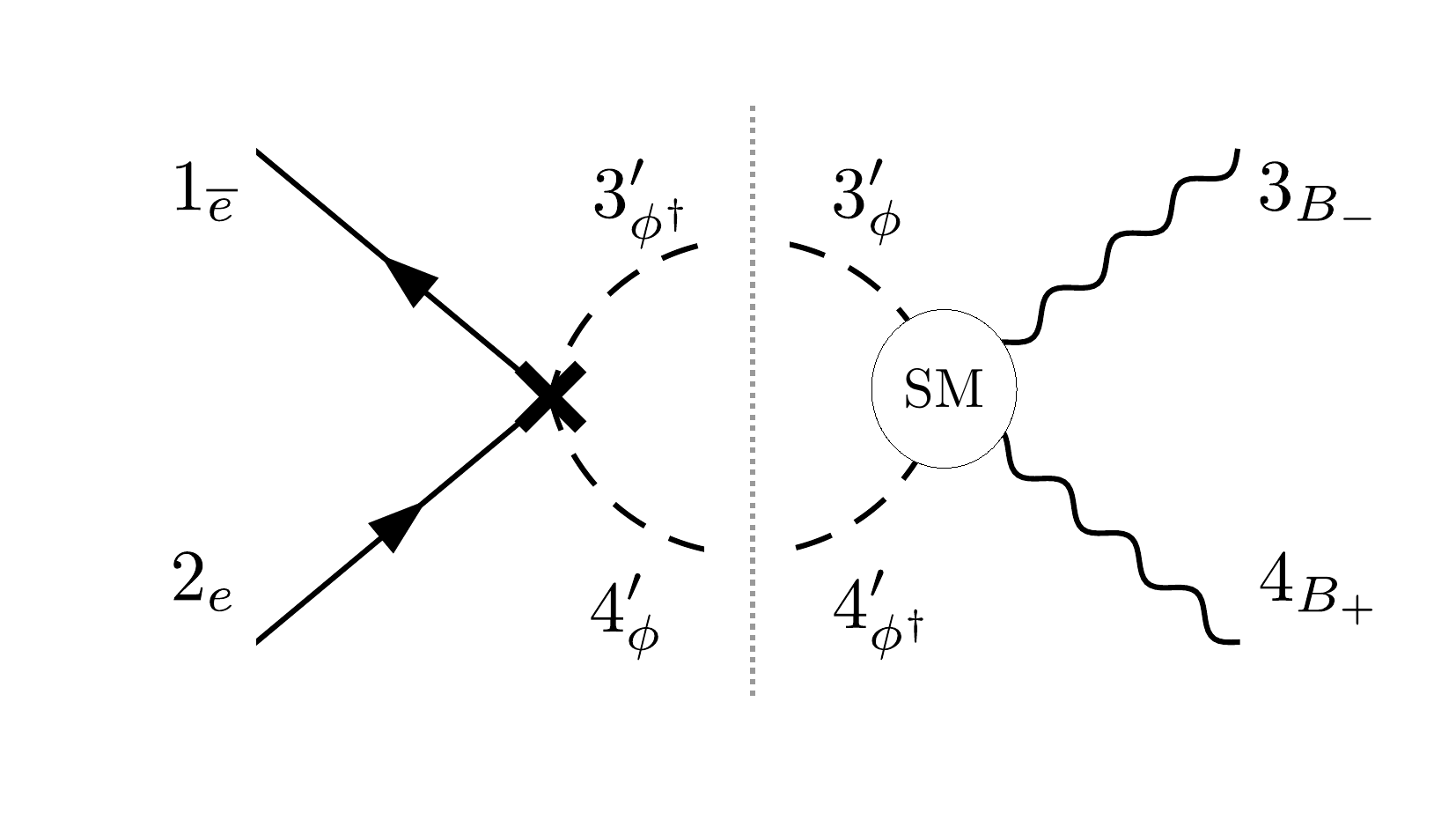}
 \end{center}
 \caption{\it Double cut for the mixing of $e^2\phi^2 D^3$ into $e^2 B^2 D$. The cross indicates the EFT (minimal) amplitude, while the bubble stands for the SM (non-necessarily minimal) amplitude.}\label{fig:cuts} 
\end{figure}

This result is not inmediate from the on-shell amplitude approach either. To see why, let us express the aforementioned operators in the amplitude basis:
\begin{align}
 \widetilde{\mathcal{A}}_{e^2\phi^2 D^3}^{(1)} (1_{\overline{e}},2_{e},3_{\phi^\dagger},4_{\phi}) &= \langle 12\rangle \langle 13\rangle [12][23]\,,\\
 \widetilde{\mathcal{A}}_{e^2\phi^2 D^3}^{(2)}(1_{\overline{e}},2_{e},3_{\phi^\dagger},4_{\phi}) &= \langle 13\rangle \langle 23\rangle [23]^2\,,\\
 \mathcal{A}_{e^2B^2D} (1_{\overline{e}}, 2_e, 3_{B_{-}}, 4_{B_{+}}) &= \langle 13 \rangle^2 [14] [24]\,.
\end{align}
We can compute the mixing of the first two amplitudes into the second performing the usual unitarity double cuts~\cite{Caron-Huot:2016cwu,EliasMiro:2020tdv,Baratella:2020lzz,AccettulliHuber:2021uoa}; see Fig.~\ref{fig:cuts}. The SM amplitude in the cut reads:
\begin{equation}
 \mathcal{A}_\text{SM}(1_\phi, 2_{\phi^\dagger}, 3_{B_-}, 4_{B_+}) = \frac{1}{2} g_1^2 \frac{\langle 13\rangle \langle 23\rangle}{\langle 14\rangle \langle 24\rangle}\,.
\end{equation}
Thus, we have:
\begin{equation}
 \gamma_{c_{e^2B^2D},\,\tilde{c}_{e^2\phi^2 D^3}^{(1)}} \mathcal{A}_{e^2B^2D}(1_{\overline{e}}, 2_e, 3_{B_{-}}, 4_{B_{+}}) = -\frac{4}{\pi} \mathcal{I}_1\,,
\end{equation}
with
\begin{align}
 \mathcal{I}_1 &=  \int d\text{LIPS} \, \widetilde{\mathcal{A}}_{e^2\phi^2 D^3}^{(1)} (1_{\overline{e}},2_{e},3_{\phi^\dagger},4_{\phi}) \times \mathcal{A}_\text{SM}(-3'_\phi, -4'_{\phi^\dagger} ,3_{B_-}, 4_{B_+})\nonumber\\
 &= \frac{1}{2} g_1^2\int d\text{LIPS} \, \langle 12\rangle \langle 13'\rangle [12][23'] \, \frac{\langle 3'3\rangle \langle 4'3\rangle}{\langle 3'4\rangle \langle 4'4\rangle}\nonumber\\
 &=-\frac{1}{4}g_1^2\int_0^{\pi/2}\int_0^{2\pi}d\varphi\, d\theta\, s_\theta\, c_\theta\, e^{\ii\varphi}\, \langle 12\rangle [12] \left(\mc{ c_\theta \langle 13\rangle-e^{\ii\varphi}s_\theta \langle 14\rangle}\right) \left(e^{\ii\varphi}c_\theta[23]-s_\theta[24]\right)
 &= 0\,, 
\end{align}
which implies
\begin{equation}\label{eq:explicit1}
 \gamma_{c_{e^2B^2D},\,\tilde{c}_{e^2\phi^2 D^3}^{(1)}} = 0\,.
\end{equation}
Note that, contrary to those zeros highlighted by non-renormalisation theorems~\cite{Cheung:2015aba,Craig:2019wmo}, in this case the SM amplitude does not vanish. The zero arises because all terms in the product of tree-level on-shell amplitudes are proportional to $e^{\ii n\varphi}$ with $n=1,2,...$, for which $\int_0^{2\pi}d\varphi e^{\ii n\varphi}$ vanishes. \mc{(As we show in Appendix~\ref{app:comparison}, this result can be reproduced though using angular-momentum selection rules~\cite{Jiang:2020rwz}.)}

The negative value of $\gamma_{c_{e^2B^2D},\,\tilde{c}_{e^2\phi^2 D^3}^{(2)}}$ seems also accidental from this perspective. In fact:
\begin{align}
 \gamma_{c_{e^2B^2D},\,\tilde{c}_{e^2\phi^2 D^3}^{(2)}} \mathcal{A}_{e^2B^2D}(1_{\overline{e}}, 2_e, 3_{B_{-}}, 4_{B_{+}}) = -\frac{4}{\pi} \mathcal{I}_2\,
\end{align}
with 
\begin{align}
 \mathcal{I}_2 &= \int d\text{LIPS} \, \widetilde{\mathcal{A}}_{e^2\phi^2 D^3}^{(2)} (1_{\overline{e}},2_{e},3_{\phi^\dagger},4_{\phi}) \times \mathcal{A}_\text{SM}(3'_\phi, 4'_{\phi^\dagger} ,3_{B_-}, 4_{B_+})\nonumber\\
 &= \frac{1}{2}g_1^2\int d\text{LIPS} \, \langle 13'\rangle \langle 23'\rangle [23']^2  \, \frac{\langle 3'3\rangle \langle 4'3\rangle}{\langle 3'4\rangle \langle 4'4\rangle}\nonumber\\
 &= -\frac{1}{2} \pi g_1^2\int_0^{\pi/2}\, d\theta \, \mc{c_\theta^3 s_\theta^3} \langle 13\rangle \langle 23\rangle [24]^2 = -\frac{1}{24}\pi g_1^2 \langle 13\rangle \langle 23\rangle [24]^2 \, , 
\end{align}
where in this case we have ommited the details about the integration over $\varphi$.

The final amplitude does not look exactly like $\mathcal{A}_{e^2 B^2 D}(1_{\overline{e}}, 2_e, 3_{B_{-}}, 4_{B_{+}})$, however:
\begin{align}
 \langle 13\rangle \langle 23\rangle [24]^2 = \langle 13\rangle \langle 32\rangle [24][42] = -\langle 13\rangle \langle 31\rangle [14][42] = -\mathcal{A}_{e^2B^2D} (1_{\overline{e}}, 2_e, 3_{B_{-}}, 4_{B_{+}})\,,
\end{align}
where in the first equality we have used the antisymmetry of angles and squares, while in the second we have used momentum conservation:
\begin{equation}
 |2\rangle\langle 2| = -|1\rangle\langle 1|-|3\rangle\langle 3|-|4\rangle\langle 4|\,.
\end{equation}
From here, we obtain that:
\begin{align} \label{eq:explicit2}
  \gamma_{c_{e^2B^2D},\,\tilde{c}_{e^2\phi^2 D^3}^{(2)}} = -\frac{g_1^2}{6}\,.
\end{align}
In the next section, we show that both the zero in Eq.~\eqref{eq:explicit1} and the negative value of $\gamma$ in Eq.~\eqref{eq:explicit2} are apparent when looking at these loops from the perspective of the positivity of two-to-two forward scattering amplitudes.

\section{Constraints from positivity}
\label{sec:positivity}
Let us consider the tree-level amplitude $\mathcal{A}$ for the elastic process $e\phi\to e\phi$ within some well-defined (local and unitary) quantum-field theory (QFT). In the forward limit, this amplitude is an analytic function of the Mandesltan invariant $s$ only, with at most single poles corresponding to particle thresholds. This permits relating the IR and UV behaviour of $\mathcal{A}$ through dispersion relations~\cite{Adams:2006sv} which, together with robust bounds on the growing of the amplitude at $s\to\infty$~\cite{Froissart:1961ux} and the optical theorem, implies that $\mathcal{A}''|_{s=0}\geq 0$. This latter quantity can be computed within the EFT, implying:
\begin{equation}\label{eq:bound}
 -\tilde{c}_{e^2\phi^2 D^3}^{(2)}\geq 0\,,
\end{equation}
while $\tilde{c}_{e^2\phi^2 D^3}^{(1)}$ remains completely unconstrained. (In the amplitude basis this result is very intuitive because the forward limit amounts to the relations $2\to 1$ and $3\to 4$, which implies that $\langle 12\rangle\langle 13\rangle [12][23]$ vanishes.)

Within this same QFT, we could compute the one-loop amplitude for $eB\to eB$, and apply an analogous reasoning~\cite{Arkani-Hamed:2020blm,Chala:2021wpj,Li:2022aby,Chala:2023jyx}, obtaining that:
\begin{equation}
 c_{e^2B^2 D}(\mu)\sim c_{e^2 B^2D}(\Lambda) + \dot{c}_{e^2 B^2 D}\log{\frac{\mu}{\Lambda}}\leq 0\Rightarrow\dot{c}_{e^2 B^2 D}\geq 0\,.
\end{equation}
In the last step we have taken into account that $e^2B^2D$ operators do not arise at tree level in weakly-coupled UV completions of the SMEFT~\cite{Craig:2019wmo}.~\footnote{\mc{If this is not the case, one can always restrict to a family of UV completions that do not generate the corresponding operator at tree level. As explained in Ref.~\cite{Chala:2023jyx}, the arguments below still apply in such case.}} Now, dimension-six tree-level interactions that might be also present in the IR do not renormalise $\mathcal{O}_{e^2 B^2 D}$, so we simply have: 
\begin{equation}\label{eq:critical1}
  \mc{\dot{c}_{e^2 B^2 D} = \alpha_1 \tilde{c}_{e^2\phi^2 D^3}^{(1)} + \alpha_2 \tilde{c}_{e^2\phi^2 D^3}^{(2)} \,,}
\end{equation}
with $\alpha_1$ and $\alpha_{2}$ some real numbers. \mc{(They are directly related to the anomalous dimensions; in fact $\alpha_1 = \gamma_{c_{e^2 B^2 D},\tilde{c}_{e^2 \phi^2 D^3}^{(1)}}$ and $\alpha_2 = \gamma_{c_{e^2 B^2 D},\tilde{c}_{e^2 
\phi^2 D^3}^{(2)}}$.)}

In order for this expression to be positive in any physical UV completion of the SMEFT, or equivalently for all values of $\tilde{c}_{e^2\phi^2 D^3}^{(1)}$ and $\tilde{c}_{e^2\phi^2 D^3}^{(1)}$ compatible with the bound in Eq.~\eqref{eq:bound}, the only possibility is that $\alpha_1=0$, $\alpha_2 = -\alpha$, with $\alpha\geq 0$:
\begin{equation}\label{eq:critical2}
 \dot{c}_{e^2 B^2 D} = (\fcolorbox{black}{LightCyan}{$\,0\,$}-\alpha) \begin{pmatrix}\tilde{c}_{e^2\phi^2 D^3}^{(1)}\tilde{c}_{e^2\phi^2 D^3}^{(2)}\end{pmatrix}\,.
\end{equation}
This is essentially Eqs.~\eqref{eq:zerofeynman}, \eqref{eq:explicit1} and \eqref{eq:explicit2}, though the explicit values of the anomalous dimensions have not been obtained. (And, of course, there has been no need for computing any loop!)

\begin{figure}[t]
 \begin{center}
 \includegraphics[width=0.6\columnwidth]{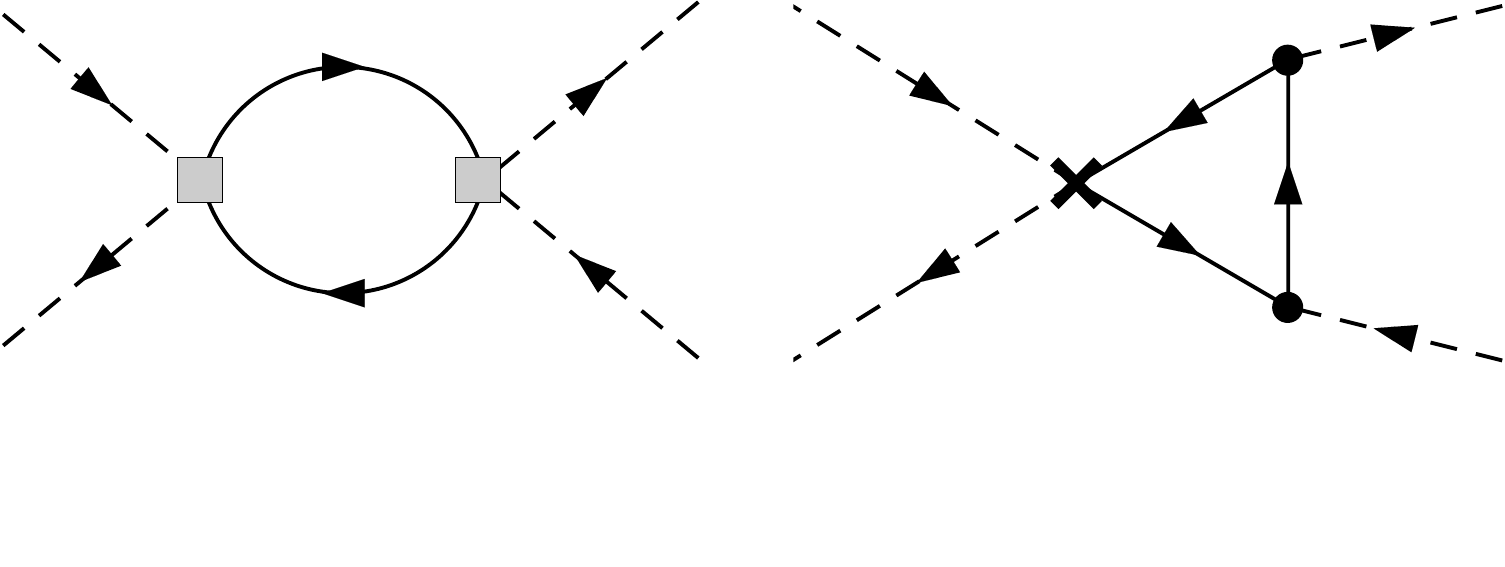}
 \end{center}
 \vspace{-1cm}
 \caption{\it Example diagrams for the mixing of $e^2 \phi^2 D^n$ operators into $\phi^4 D^4$. Vertices like \fcolorbox{black}{LightGray}{$\,\textcolor{LightGray}{i}\,$} represent dimension-six terms ($n=1$), while crosses $\boldmath{\times}$ are dimension-eight ones ($n=3$).}\label{fig:dimsix}
\end{figure}

This same reasoning can not be applied to constraining the mixing of, for example, $e^2\phi^2 D^3$ into $\phi^4 D^4$. The reason is that, in this case, one expects also the presence of dimension-six $e^2\phi^2 D$ terms which, in pairs, renormalise also $\phi^4 D^4$; see Fig.~\ref{fig:dimsix}. Hence, the positivity of $\phi^4 D^4$ can not be directly attributed to the positivity of $\gamma$ but rather to $\gamma'$~\cite{Chala:2021wpj,Chala:2023jyx}. In general, this problem arises whenever all fields in the renormalised operator are already present in the renormalising interaction. 

One further comment is in order. The crucial step from Eq.~\eqref{eq:critical1} to Eq.~\eqref{eq:critical2} relies on the conviction that $\tilde{c}_{e^2\phi^2 D^3}^{(1)}$ is not constrained at all by positivity (it can have either sign). This can be only guaranteed if the set of independent positivity restrictions is fully exhausted. For this matter, studying the elastic positivity bounds with non-superposed asymptotic states does not suffice. We must instead supplement the analysis with the scattering of states with certain degree of superposition, that we quantify as $x$; see Appendix~\ref{app:bounds}.~\footnote{Nonetheless, we have certain evidence, based on some explicit computations as well as on generalised unitarity methods, rather than on an analysis of the positivity of amplitudes in all possible UV completions of the SMEFT, that restricting to bounds valid only when $x\to 0$ is still a valid approach. This would allow us to establish more stringent constraints on the ADM of the SMEFT than those that we present in Appendix~\ref{app:adms}. We have no indisputable proof of this observation, though.}
In the case of $e\phi\to e\phi$, the bounds obtained either way are the same. In other classes of operators, however, the differences are apparent.

On the other hand, for constraining the running functions themselves, the bounds to be considered are only those that survive in the limit $x\rightarrow 0$. 
(Again, in the case at work, namely $eB\to eB$, the bounds are equivalent.) 

To explain this issue, let us imagine that we rather focus on the mixing $e^2\phi^2 D^3\to e^2 u^2 D^2$. In this case, if we take into account all independent bounds on $e^2 u^2 D^3$, we can conclude that $\dot{c}_{e^2u^2D^2}^{(1)}-3\dot{c}_{e^2u^2D^2}^{(2)}\leq 0$ and $-\dot{c}_{e^2u^2D^2}^{(1)}-\dot{c}_{e^2u^2D^2}^{(2)} \leq 0$, meaning that $c_{e^2u^2D^2}^{(1)}$ can never be the only renormalised coupling. However, an explicit computation yields $\dot{c}_{e^2u^2D^2}^{(1)} = \frac{2}{9}g_1^2 (c_{e^2\phi^2D^2}^{(2)}-c_{e^2\phi^2D^2}^{(1)})$ and $\dot{c}_{e^2u^2D^2}^{(2)}=0$ in the limit of vanishing Yukawas, \mc{in contradiction with the previous statement}. The reason is that, in the derivation of the most general positivity bounds for $e^2u^2D^2$ operators (see Eq.~\eqref{eq:e2u2full}), we find processes of the sort $ee\to ee$, which at one loop receive contributions from pairs of dimension-six terms $e^2 \phi^2 D$. The way out for ignoring the contamination from dimension-six terms is considering only those bounds on anomalous dimensions that arise in the limit $x=0$~\footnote{\mc{To see that this applies in general, take into account that the most generic positivity constraints on $\mathcal{O}_{\psi^2\psi'^2}$ operators, where $\psi,\psi'$ are arbitrary fields, ensue from processes involving superposition states; see Appendix~\ref{app:bounds}. This implies the appereance of $\psi^{(')}\psi^{(')}\to\psi^{(')}\psi^{(')}$ processes on top of the standard $\psi\psi'\to\psi\psi'$. Even restricting to UV completions that do not generate $\mathcal{O}_{\psi^2\psi'^2}$ at tree level, the first two processes receive contributions from pairs of dimension-six $\mathcal{O}_{\psi^2\psi'^2}$ with two $\psi$ (or $\psi'$) in a loop.}}.

\section{Structure of the anomalous dimension matrix}
\label{sec:structure}
\begin{figure}[t]
\begin{center}
  \includegraphics[width=0.68\columnwidth]{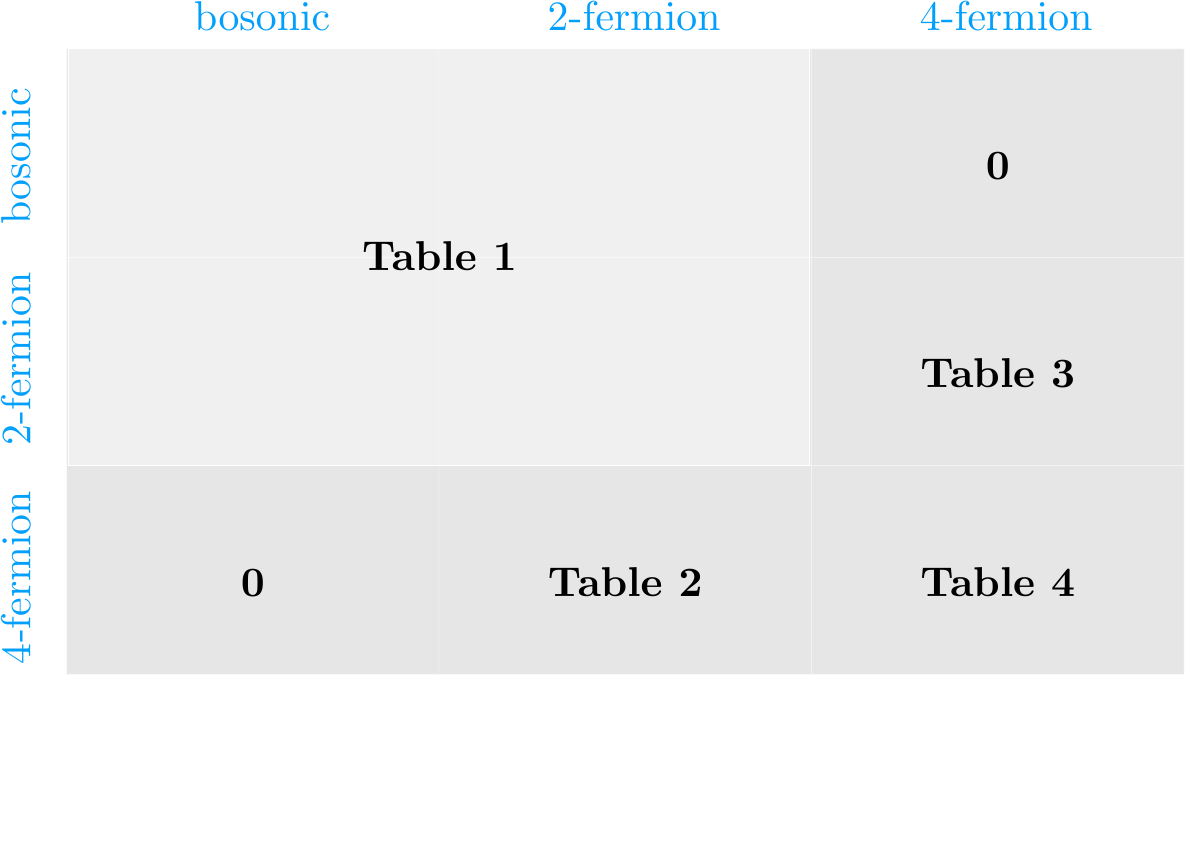}
\end{center}
\vspace{-1.5cm}
\caption{\it Schematic structure of the SMEFT ADM. The different blocks are given in Tabs.~\ref{tab:adm1}, ~\ref{tab:adm3}, ~\ref{tab:adm2} and ~\ref{tab:adm4}}\label{fig:rges}
\end{figure}

In light of the previous discussion, we aim to constrain anomalous dimensions $\gamma_{ij}$ of four-field interactions, where $i$ represents a Wilson coefficient that is bounded in the limit $x=0$, and $j$ is any other coefficient of a tree-level operator.

To this aim, we choose a basis in which all $x$-independent positivity bounds are decoupled. This amounts to picking one (in general there might be many) rotation matrix $R$  such that $P R=\mathbb{I}_{P}$, where $P$ is the matrix of $x$-independent positivity bounds, namely that satisfying $P\vec{c}\geq 0$; and $\mathbb{I}_P$ is the matrix consisting in as many rows of the identity matrix as number of positivity bounds.

For example, we can have:
\begin{align}
 P_{u^2\phi^2D^3} &= \begin{pmatrix}-1&-1\end{pmatrix}\,,\quad R_{u^2\phi^2D^3} = \begin{pmatrix}2&3\\-3&-3\end{pmatrix}\,,\\[0.2cm]
 P_{l^2u^2D^2} &= \begin{pmatrix}\mc{0}&\mc{-1}\end{pmatrix}\,,\quad\quad R_{l^2u^2D^2} = \begin{pmatrix}2&3\\-1&0\end{pmatrix}\,.
\end{align}
In this new basis, in which we denote the Wilson coefficients with a tilde, $\tilde{c}$, the $x$-independent bound for $u^2\phi^2D^2$ reads simply $\tilde{c}_{u^2\phi^2D^3}^{(1)}\geq 0$. The bound for $l^2u^2 D^2$ valid in the limit $x\to 0$ is $\tilde{c}_{l^2u^2D^2}^{(1)}\geq 0$. From here, it follows that
\begin{align}
 0\geq \dot{c}_{l^2u^2D^2}^{(1)} &= -\alpha \tilde{c}_{u^2\phi^2 D^3}^{(1)} + \cdots 
\end{align}
with $\alpha\geq 0$ and where the ellipses represent operators of other classes which are completely independent. Hence, $\gamma_{\tilde{c}_{l^2 u^2D^2}\,,\, \tilde{c}_{u^2\phi^2 D^3}^{(1)}} \leq 0$ while $\gamma_{\tilde{c}_{l^2u^2 D^2}\,,\,\tilde{c}_{u^2\phi^2 D^3}^{(2)}}=0$. 

The process of obtaining a full colection of rotation matrices can be easily automatised. We provide one in the ancillary file \texttt{matrices.txt}.
Some of them are completely determined by the requirement $PR=\mathbb{I}_P$; for the rest we have simply fixed one of the many possibilities that fulfill $\text{Det}(R)=1$. (In particular, the matrix $P_{e^2\phi^2 D^3}$ is different from that shown in Eq.~\eqref{eq:basistransformation}.) 

The reader will notice that in certain classes of operators, for instance in $G^2\phi^2 D^2$, the matrices are smaller then the number of operators. This is simply because we ignore all those Wilson coefficients beyond the last one intervening in the positivity bounds in Appendix~\ref{app:bounds}, such as for example $c_{G^2\phi^2 D^2}^{(3)}$. 

In the basis defined by these rotations, the different blocks of the SMEFT ADM (see Fig.~\ref{fig:rges}) are described by Tabs.~\ref{tab:adm1}, ~\ref{tab:adm2}, ~\ref{tab:adm3} and ~\ref{tab:adm4} in Appendix~\ref{app:adms}. It should be noticed that operators not constrained by positivity are not shown in the left rows, while operators that only appear at one-loop in UV completions of the SMEFT are absent from the top columns. Besides, we provide cross-checks of some of these entries in Appendix~\ref{app:xchecks}.

\section{Discussion and outlook}
\label{sec:conclusions}
Even if two different classes of dimension-8 SMEFT operators mix under renormalisation group running at one-loop, there are always certain interactions in one class that do not mix into certain other interactions in the second class, or that they do it in such a way that the running has always definite sign, irrespective of the SM couplings. In the appropriate basis of operators, these effects manifest as zeros or definite-sign entries in the ADM.

We have shown that such a basis, or equivalently the corresponding rotation matrix $\mathcal{R}$ that must be applied to the commonly used basis of Ref.~\cite{Murphy:2020rsh}, can not be easily found without computing explicitly the one-loop RGEs, either using Feynman diagrams or on-shell amplitude methods. However, following the results of Ref.~\cite{Chala:2023jyx}, we have derived $\mathcal{R}$ basing solely on positivity bounds obtained from tree-level computations in the forward limit. These bounds, which extend those previously computed in the literature~\cite{Remmen:2019cyz,Bi:2019phv,Remmen:2020vts,Trott:2020ebl,Yamashita:2020gtt,Gu:2020ldn,Li:2021lpe,Li:2022tcz}, particularly in the   $G^2\phi^2D^2$ and $X^2\psi^2D$ sectors, are of major importance for constraining the SMEFT parameter space. Hence, our work adds also to this aspect of the SMEFT phenomenology.

In the new basis defined by $\cal{R}$, we highlight 84 non-trivial zeros and much more negative entries in the ADM. We emphasise that, following our same methods, different bases involving different number of zeros and negative entries can be obtained; in this paper we have simply singled out one for the sake of example.  It would be actually interesting to know which is such basis in which the number of zeros is maximised. The main point is that we can do so without performing any involved one-loop calculation. It would be also important to cross-check these results by explicit computations, some of which we provide in Appendix~\ref{app:xchecks}, and eventually complete the renormalisation of the dimension-8 SMEFT.

\section*{Acknowledgments}
We thank Jose Santiago for useful discussions.
MC is supported by the Ram\'on y Cajal grant  RYC2019-027155-I funded by the MCIN /AEI/10.13039/501100011033 and by ``El FSE invierte en tu futuro'', and acknowledges support from the grant  CNS2022-136024 funded by the European Union NextGenerationEU/PRTR, from grants PID2021-128396NB-I00 and PID2022-139466NB-C22 funded by ``FEDER Una manera de hacer Europa'' as well as from the Junta de Andalucía grants FQM 101 and P21-00199. XL is supported in part by the National Natural Science Foundation of China under grant No. 11835013.

\appendix

\section{Comparison with Ref.~\cite{Jiang:2020rwz}}
\label{app:comparison}
It has been proven~\cite{Jiang:2020rwz} that angular-momentum conservation forbids the mixing between any two operators with $n$ common legs $\mathcal{I}$ if they differ on the value of $j(j+1)$, defined up to a factor as the eigenvalue of the operator $W_\mathcal{I}$ that acts on an amplitude $\mathcal{A}$ as:
\begin{equation}
W_\mathcal{I}^2(\mathcal{A}) =\frac{1}{8}P_\mathcal{I}^2 \bigg[\text{Tr}(M_\mathcal{I}^2 \mathcal{A}) + \text{Tr}(\tilde{M}_\mathcal{I}^2 \mathcal{A})\bigg] - \frac{1}{4} P_\mathcal{I}^{\dot{a}a} P_\mathcal{I}^{\dot{b}b} (M_{\mathcal{I}ab} \tilde{M}_{\mathcal{I}\dot{a}\dot{b}} \mathcal{A})\,,
\end{equation}
where
\begin{align}
 M_{\mathcal{I}ab} &= \ii \sum_{i\in\mathcal{I}}\left(\lambda_{ia}\frac{\partial}{\partial\lambda_i^b}+\lambda_{ib}\frac{\partial}{\partial\lambda_i^a}\right)\,,\\
 \tilde{M}_{\mathcal{I}\dot{a}\dot{b}} &= \ii \sum_{i\in\mathcal{I}}\left(\tilde{\lambda}_{i\dot{a}}\frac{\partial}{\partial\tilde{\lambda}_i^{\dot{b}}}+\tilde{\lambda}_{i\dot{b}}\frac{\partial}{\partial\tilde{\lambda}_i^{\dot{a}}}\right)\,;
\end{align}
while $M^2_{\mathcal{I}ab}\mathcal{A} = M_{\mathcal{I}a}^c M_{\mathcal{I}cb}\mathcal{A}$ and $P_\mathcal{I} = \sum_{i\in\mathcal{I}}\lambda_i\tilde{\lambda}_i$.

This explains our finding in Eq.~\eqref{eq:explicit1}. Indeed, let us first consider the amplitude $\mathcal{A}_{e^2B^2 D} = \langle 13 \rangle^2 [14] [24]$ in the channel $\mathcal{I} = \lbrace 1,2\rbrace$. With the help of \texttt{SpinorHelicity4D}~\cite{AccettulliHuber:2023ldr}, we obtain:
\begin{align}
 M_{\mathcal{I}ad}(\mathcal{A}_{e^2 B^2 D}) &= 2 \ii \langle 13\rangle^2 \langle 14\rangle [24] (\lambda_{1d} \lambda_{3a}+\lambda_{1a}\lambda_{3d})\,,\\
 \tilde{M}_{\mathcal{I}\dot{a}\dot{d}}(\mathcal{A}_{e^2 B^2 D}) &= -\ii\langle 13\rangle^2 \left[\left([24]\tilde{\lambda}_{1\dot{d}}+[14]\tilde{\lambda}_{2\dot{d}}\right)\tilde{\lambda}_{4\dot{a}}+\left([24]\tilde{\lambda}_{1\dot{a}}+[14]\tilde{\lambda}_{2\dot{a}}\right)\tilde{\lambda}_{4\dot{d}}\right]\,,
\end{align}
from where
\begin{align}
 W_\mathcal{I}^2(\mathcal{A}_{e^2B^2 D}) &=
 2 s(\langle 13\rangle \langle 23\rangle [24]^2-2 \mathcal{A}_{e^2B^2D})\nonumber\\
 &= -6 s \langle 13\rangle \langle 23\rangle [24]^2\,\,,
\end{align}
which indicates that $j=2$.

On the contrary, for $\widetilde{\mathcal{A}}_{e^2\phi^2D^3}^{(1)}= \langle 12\rangle \langle 13\rangle [12][23]$, we find:
\begin{align}
 W_\mathcal{I}^2(\widetilde{\mathcal{A}}_{e^2\phi^2D^3}^{(1)}) &= -2 s \widetilde{\mathcal{A}}_{e^2\phi^2 D^3}^{(1)}
\end{align}
and so it has $j=1$. Because both values of $j$ differ, the first operator can not mix into the other.

Meanwhile, 
\begin{equation}
W^2_\mathcal{I} (\widetilde{\mathcal{A}}_{e^2\phi^2 D^3}^{(2)}) = -2s\widetilde{\mathcal{A}}_{e^2\phi^2 D^3}^{(1)}-6s\widetilde{\mathcal{A}}_{e^2\phi^2 D^3}^{(2)}\,,
\end{equation}
which is a linear combination of $j=1$ and $j=2$, and hence the mixing of this amplitude into $\mathcal{A}_{e^2 B^2 D}$ is, as we found, allowed.

Several comments are in order though. \textit{(i)} Despite being possible to automatise, this calculation is long. In contrast, in this same amplitude basis, the positivity constraint is apparent: $\langle 12\rangle \langle 13\rangle [12][23]\to \langle 11\rangle \langle 13\rangle [11][13]$ in the forward limit, and so it vanishes, while $\langle 13\rangle^2 [14][24]\to \langle 13\rangle^2 [13]^2$ does not, so the first can not mix into the second.
\textit{(ii)} While more widely applicable, it is far from evident whether all the results we obtain here can be reproduced using the approach of Ref.~\cite{Jiang:2020rwz}.
\textit{(iii)} Most importantly, to the best of our knowledge, the angular-momentum method does not provide information about the signs of anomalous dimensions. 

In light of all this, the most we can say is that the methods of Ref.~\cite{Jiang:2020rwz}, based on angular momentum conservation, and those applied here, relying on positivity, complement each other; each one being preferred depending on the context.

\section{Positivity bounds}
\label{app:bounds}
The most stringent positivity bounds can be derived from the elastic scattering of two superposition states $|u\rangle=\sum_i u_i |i \rangle$ and  $|v\rangle=\sum_j v_j |j \rangle$:
\begin{equation}
 \mathcal{A}(|u\rangle|v\rangle  \to |u\rangle|v\rangle)=\sum_{ijkl} u_i v_j u_k^* u_l^*\mathcal{A}(|i\rangle|j\rangle\to |i\rangle|j\rangle)\,.
\end{equation}
Assuming $u_i$ and $v_j$ to be real, and adopting the self-conjugate particle basis, this amplitude in the forward limit fulfills~\cite{Bellazzini:2015cra,Cheung:2016yqr,deRham:2018qqo}
\begin{align}
 \mathcal{A}^{uvuv}\equiv \frac{d^2}{ds^2}\mathcal{A}(|u\rangle|v\rangle  \to |u\rangle|v\rangle) = \sum_{ijkl}u_i v_j u_k v_l  \mathcal{A}^{ijkl}\geq 0\,.
\end{align}

In practice, we parameterise the superposition via a quantity $x$. For example, for constraining operators of the type $B^2\phi^2 D^2$ we consider $|u\rangle = |\phi\rangle+x|B\rangle$ as well as $|v\rangle = |\phi\rangle+|B\rangle$. 

Below, for each class of operators, we quote the set of independent positivity bounds valid in general (namely for any superposition), as well as those valid only for $x=0$, which we show in gray. (When both coincide, we show only one.)

\subsection{$\phi^4D^4$}
\bea
c_{\phi^4D^4}^{\left(2\right)}\geq0\,,\quad c_{\phi^4D^4}^{\left(2\right)}+c_{\phi^4D^4}^{\left(2\right)}\geq0\,,\quad c_{\phi^4D^4}^{\left(1\right)}+c_{\phi^4D^4}^{\left(2\right)}+c_{\phi^4D^4}^{\left(3\right)}\geq0 \,.
\eea
\subsection{$G^2\phi^2D^2$}
\begin{align}
 -c_{G^2\phi^2D^2}^{\left(1\right)}-2c_{G^2\phi^2D^2}^{\left(2\right)}\geq0\,,\ \ c_{G^2\phi^2D^2}^{\left(2\right)}\geq0\,,\\
 \textcolor{gray}{-c_{G^2\phi^2D^2}^{\left(1\right)}\geq0}\,.
\end{align}

\subsection{$W^2\phi^2D^2$}
\bea
-c_{W^2\phi^2D^2}^{\left(1\right)}-2c_{W^2\phi^2D^2}^{\left(2\right)}\geq0 \,, \quad c_{W^2\phi^2D^2}^{\left(2\right)}\geq 0 \,, \nonumber\\ -c_{W^2\phi^2D^2}^{\left(1\right)} + c_{W^2\phi^2D^2}^{\left(4\right)} \geq 0\,, \quad   -c_{W^2\phi^2D^2}^{\left(1\right)} - c_{W^2\phi^2D^2}^{\left(4\right)} \geq 0 \,, \\
\textcolor{gray}{- c_{W^2\phi^2D^2}^{\left(1\right)} \geq0  \;.}
\eea
\subsection{$B^2\phi^2D^2$}
\bea
-c_{B^2\phi^2D^2}^{\left(1\right)}-2c_{B^2\phi^2D^2}^{\left(2\right)}\geq0,\ \ c_{B^2\phi^2D^2}^{\left(2\right)}\geq0 \,,\\
\textcolor{gray}{-c_{B^2\phi^2D^2}^{\left(1\right)}\geq0 \;.}
\eea
\subsection{$l^2\phi^2D^3$}
\bea
- c_{l^2\phi^2D^3}^{\left(1\right)}-c_{l^2\phi^2D^3}^{\left(2\right)}-c_{l^2\phi^2D^3}^{\left(3\right)}-c_{l^2\phi^2D^3}^{\left(4\right)}\geq0 \,, \nonumber \\  - c_{l^2\phi^2D^3}^{\left(1\right)}-c_{l^2\phi^2D^3}^{\left(2\right)}+c_{l^2\phi^2D^3}^{\left(3\right)}+c_{l^2\phi^2D^3}^{\left(4\right)}\geq0 \;.
\eea
\subsection{$e^2\phi^2D^3$}
\bea
-c_{{e^2\phi}^2D^3}^{\left(1\right)}-c_{{e^2\phi}^2D^3}^{\left(2\right)}\geq0\,.
\eea
\subsection{$q^2\phi^2D^3$}
\bea
&& -c_{q^2\phi^2D^3}^{\left(1\right)}-c_{q^2\phi^2D^3}^{\left(2\right)}+c_{q^2\phi^2D^3}^{\left(3\right)}+c_{q^2\phi^2D^3}^{\left(4\right)}\geq0,  \nonumber \\
&& -c_{q^2\phi^2D^3}^{\left(1\right)}-c_{q^2\phi^2D^3}^{\left(2\right)}-c_{q^2\phi^2D^3}^{\left(3\right)}-c_{q^2\phi^2D^3}^{\left(4\right)}\geq0 \;. 
\eea
\subsection{$u^2\phi^2D^3$}
\bea
 -c_{u^2\phi^2D^3}^{\left(1\right)}-c_{u^2\phi^2D^3}^{\left(2\right)}\geq0\,. 
  \eea
\subsection{$d^2\phi^2D^3$}
\bea
  -c_{d^2\phi^2D^3}^{\left(1\right)}-c_{d^2\phi^2D^3}^{\left(2\right)}\geq0\,. 
  \eea
\subsection{$q^2G^2D$}
\bea
 -3c_{q^2G^2D}^{\left(1\right)}-c_{q^2G^2D}^{(3)}\geq0 \,,\quad -3c_{q^2G^2D}^{\left(1\right)}+c_{q^2G^2D}^{(3)}\geq0 \;. 
  \eea
\subsection{$q^2W^2D$ } 
\bea
-c_{q^2W^2D}^{\left(1\right)}\ \geq0 \;.
\eea
\subsection{$q^2B^2D$ }
\bea
-c_{q^2B^2D}^{\left(1\right)}\ \geq0 \,.
\eea
\subsection{$u^2G^2D$}
\bea
 -3c_{u^2G^2D}^{(1)}-c_{u^2G^2D}^{(3)}\geq0,\quad
-3c_{u^2G^2D}^{\left(1\right)}+c_{u^2G^2D}^{(3)}\geq0 \;. 
    \eea
\subsection{$u^2W^2D$  }
\bea
-c_{u^2W^2D} \geq0 \,.
\eea
\subsection{$u^2B^2D$  }
\bea
-c_{u^2B^2D} \geq0 \,.
\eea
\subsection{$d^2G^2D$}
\bea
 -3c_{d^2G^2D}^{(1)}-c_{d^2G^2D}^{(3)}\geq0\,,\quad
-3c_{d^2G^2D}^{\left(1\right)}+c_{d^2G^2D}^{(3)}\geq0 \,. 
    \eea
\subsection{$d^2W^2D$}
\bea
-c_{d^2W^2D} \geq0\,.
\eea
\subsection{$d^2B^2D$}
\bea
-c_{d^2B^2D} \geq0\,.
\eea
\subsection{$l^2G^2D$}
\bea
  -c_{l^2G^2D}^{} \geq0 \,.  
    \eea
\subsection{$l^2W^2D$}
\bea
-c_{l^2W^2D}^{\left(1\right)}\geq0\,.
\eea
\subsection{$l^2B^2D$}
\bea
-c_{l^2B^2D}\geq0\,.
\eea
\subsection{$e^2G^2D$}
\bea
-c_{e^2G^2D}^{} \geq0 \,. 
\eea
\subsection{$e^2W^2D$}
\bea
-c_{e^2W^2D}\geq0\,.
\eea
\subsection{$e^2B^2D$}
\bea
-c_{e^2B^2D}\geq0\,.
\eea
\subsection{$l^4D^2$}
\bea
-c_{l^4D^2}^{\left(2\right)}\geq0,\ \ -c_{l^4D^2}^{\left(1\right)}-  3 c_{l^4D^2}^{\left(2\right)}\geq0\,.
\eea
\subsection{$q^4 D^2$}
\bea
    -c_{q^4D^2}^{\left(1\right)}-3c_{q^4D^2}^{\left(2\right)}-c_{q^4D^2}^{\left(3\right)}-3c_{q^4D^2}^{\left(4\right)}&\geq&0\,, \nonumber \\
 -c_{q^4D^2}^{\left(2\right)}-c_{q^4D^2}^{\left(3\right)}&\geq&0 \,,  \nonumber \\
 -c_{q^4D^2}^{\left(2\right)}-c_{q^4D^2}^{\left(4\right)}&\geq&0,\nonumber \\
 -c_{q^4D^2}^{\left(2\right)}+c_{q^4D^2}^{\left(4\right)}&\geq&0 \,.
\eea
\subsection{$l^2q^2D^2$}
\bea
  c_{l^2q^2D^2}^{\left(1\right)}-c_{l^2q^2D^2}^{\left(2\right)}+c_{l^2q^2D^2}^{\left(3\right)}-c_{l^2q^2D^2}^{\left(4\right)}&\geq& 0\,,\nonumber \\
  c_{l^2q^2D^2}^{\left(1\right)}-c_{l^2q^2D^2}^{\left(2\right)}-c_{l^2q^2D^2}^{\left(3\right)}+c_{l^2q^2D^2}^{\left(4\right)} &\geq&0\,,\nonumber \\
-c_{l^2q^2D^2}^{\left(1\right)}-3c_{l^2q^2D^2}^{\left(2\right)}-c_{l^2q^2D^2}^{\left(3\right)}-3c_{l^2q^2D^2}^{\left(4\right)}&\geq&0\,,\nonumber \\
-c_{l^2q^2D^2}^{\left(1\right)}-3c_{l^2q^2D^2}^{\left(2\right)}+c_{l^2q^2D^2}^{\left(3\right)}+3c_{l^2q^2D^2}^{\left(4\right)}&\geq& 0\,,\nonumber \\
-c_{l^2q^2D^2}^{\left(2\right)}+c_{l^2q^2D^2}^{\left(3\right)}+2c_{l^2q^2D^2}^{\left(4\right)}&\geq& 0\,,\nonumber \\
-c_{l^2q^2D^2}^{\left(2\right)}-c_{l^2q^2D^2}^{\left(3\right)}&\geq& 0\,,  \\
\textcolor{gray}{
-c_{l^2q^2D^2}^{\left(2\right)}\ -c_{l^2q^2D^2}^{\left(4\right)}\geq0 \,,\quad
\ -c_{l^2q^2D^2}^{\left(2\right)}\ +c_{l^2q^2D^2}^{\left(4\right)}} &\textcolor{gray}{ \geq } &\textcolor{gray}{ 0\,.}
\eea
\subsection{$e^4D^2$}
\bea
-c_{e^4D^2}\geq0\,.
\eea
\subsection{$u^4 D^2$}
\bea
 { -c_{u^4D^2}^{\left(1\right)}-3c_{u^4D^2}^{\left(2\right)}\geq0,\quad -c_{u^4D^2}^{\left(2\right)}\geq0 \,.}
    \eea
\subsection{$d^4 D^2$}
\bea
 {-c_{d^4D^2}^{\left(1\right)}-3c_{d^4D^2}^{\left(2\right)}\geq0\,,\quad -c_{d^4D^2}^{\left(2\right)}\geq0 \,.}
    \eea
\subsection{$e^2u^2D^2$}
\bea
{-c_{e^2u^2D^2}^{\left(1\right)}-3c_{e^2u^2D^2}^{\left(2\right)}\geq0\,,\quad c_{e^2u^2D^2}^{\left(1\right)}-c_{e^2u^2D^2}^{\left(2\right)}\geq0 \,,}\label{eq:e2u2full} \\
\textcolor{gray}{-c_{e^2u^2D^2}^{\left(2\right)}\geq0\;.}\label{eq:e2u2limit}
\eea
\subsection{$e^2d^2D^2$}
\bea
{-c_{e^2d^2D^2}^{\left(1\right)}-3c_{e^2d^2D^2}^{\left(2\right)}\geq0\,,\quad c_{e^2d^2D^2}^{\left(1\right)}-c_{e^2d^2D^2}^{\left(2\right)}\geq0 \,,} \\
\textcolor{gray}{-c_{e^2d^2D^2}^{\left(2\right)}\geq0\;.}
\eea
\subsection{$u^2d^2D^2$  }
\bea
 -3c_{u^2d^2D^2}^{\left(1\right)}-9c_{u^2d^2D^2}^{\left(2\right)}-c_{u^2d^2D^2}^{\left(3\right)}-3c_{u^2d^2D^2}^{\left(4\right)}&\geq&0 \,,\nonumber \\ 
   3c_{u^2d^2D^2}^{\left(1\right)}-3c_{u^2d^2D^2}^{\left(2\right)}+c_{u^2d^2D^2}^{\left(3\right)}-c_{u^2d^2D^2}^{\left(4\right)}&\geq&0 \,, \nonumber \\ 
-12c_{u^2d^2D^2}^{\left(2\right)}+3c_{u^2d^2D^2}^{\left(3\right)}+5c_{u^2d^2D^2}^{\left(4\right)}&\geq&0 \,, \nonumber \\ 
-12c_{u^2d^2D^2}^{\left(2\right)}-3c_{u^2d^2D^2}^{\left(3\right)}-c_{u^2d^2D^2}^{\left(4\right)}&\geq& 0 \;. \\
\textcolor{gray}{-3c_{u^2d^2D^2}^{\left(2\right)}\ -c_{u^2d^2D^2}^{\left(4\right)}\geq0 \,, \quad
-6c_{u^2d^2D^2}^{\left(2\right)}+c_{u^2d^2D^2}^{\left(4\right)}} &\textcolor{gray}{\geq} & \textcolor{gray}{ 0\;.}
    \eea
\subsection{$l^2e^2D^2$}
\bea
-c_{l^2e^2D^2}^{\left(1\right)}-c_{l^2e^2D^2}^{\left(2\right)}\geq0\,,\ \ \ c_{l^2e^2D^2}^{\left(1\right)}-3c_{l^2e^2D^2}^{\left(2\right)}\geq0 \,, \\
\textcolor{gray}{-c_{l^2e^2D^2}^{\left(2\right)}\geq0 \;.}
\eea
\subsection{$l^2u^2D^2$}
\bea
c_{l^2u^2D^2}^{\left(1\right)}-3c_{l^2u^2D^2}^{\left(2\right)}\geq0 \,,\ \ -c_{l^2u^2D^2}^{\left(1\right)}-c_{l^2u^2D^2}^{\left(2\right)}\geq0\,, \\
\textcolor{gray}{-c_{l^2u^2D^2}^{\left(2\right)}\geq0\,,}
    \eea
\subsection{$l^2d^2D^2$}
\bea
c_{l^2d^2D^2}^{\left(1\right)}-3c_{l^2d^2D^2}^{\left(2\right)}\geq0\,,\ \ -c_{l^2d^2D^2}^{\left(1\right)}-c_{l^2d^2D^2}^{\left(2\right)}\geq0 \,, \\
\textcolor{gray}{ -c_{l^2d^2D^2}^{\left(2\right)}\geq0\;.}
\eea
\subsection{$q^2e^2D^2$}
\bea
 c_{q^2e^2D^2}^{\left(1\right)}-3c_{q^2e^2D^2}^{\left(2\right)}\geq0\,,\ \ -c_{q^2e^2D^2}^{\left(1\right)}-c_{q^2e^2D^2}^{\left(2\right)}\geq0 \,, \\
\textcolor{gray}{
-c_{q^2e^2D^2}^{\left(2\right)}\geq0\;.}
\eea
\subsection{$q^2u^2D^2$}
\bea
 3c_{q^2u^2D^2}^{\left(1\right)}-9c_{q^2u^2D^2}^{\left(2\right)}+c_{q^2u^2D^2}^{\left(3\right)}-3c_{q^2u^2D^2}^{\left(4\right)}&\geq& 0 \,, \nonumber \\
-3c_{q^2u^2D^2}^{\left(1\right)}-3c_{q^2u^2D^2}^{\left(2\right)}-c_{q^2u^2D^2}^{\left(3\right)}-c_{q^2u^2D^2}^{\left(4\right)}&\geq&0 \,,\nonumber \\
-6c_{q^2u^2D^2}^{\left(2\right)}+c_{q^2u^2D^2}^{\left(4\right)}&\geq&0 \,, \\
\textcolor{gray}{
-3c_{q^2u^2D^2}^{\left(2\right)}\ -c_{q^2u^2D^2}^{\left(4\right)}\geq0\,, \quad
-6c_{q^2u^2D^2}^{\left(2\right)}+c_{q^2u^2D^2}^{\left(4\right)}} & \textcolor{gray}{\geq} & \textcolor{gray}{0\;.
}\eea
\subsection{$q^2d^2D^2$}
\bea
 3c_{q^2d^2D^2}^{\left(1\right)}-9c_{q^2d^2D^2}^{\left(2\right)}+c_{q^2d^2D^2}^{\left(3\right)}-3c_{q^2d^2D^2}^{\left(4\right)}&\geq& 0 \,, \nonumber \\
 -3c_{q^2d^2D^2}^{\left(1\right)}-3c_{q^2d^2D^2}^{\left(2\right)}-c_{q^2d^2D^2}^{\left(3\right)}-c_{q^2d^2D^2}^{\left(4\right)}&\geq& 0 \,, \nonumber \\
-6c_{q^2d^2D^2}^{\left(2\right)}+c_{q^2d^2D^2}^{\left(4\right)}&\geq& 0 \,, \\
\textcolor{gray}{
-3c_{q^2d^2D^2}^{\left(2\right)}-c_{q^2d^2D^2}^{\left(4\right)}\geq0\,, \quad
-6c_{q^2d^2D^2}^{\left(2\right)}+c_{q^2d^2D^2}^{\left(4\right)}} & \textcolor{gray}{\geq} & \textcolor{gray}{0 \;. }
\eea

\section{Blocks of the anomalous dimension matrix}
\label{app:adms}

In the following tables, we represent with a blue minus sign (\textcolor{blue}{$-$}) ADM entries that are neccessarily negative in the basis provided in the ancillary file \texttt{matrices.txt}. Trivial zeros, arising for example in cases in which the renormalising and the renormalised operators have no fields in common, are specified with a simple zero ($0$).  On the contrary, non-trivial zeros are represented with a blue-boxed zero (\fcolorbox{black}{LightCyan}{$\,0\,$}).

ADM elements that we can not constrain from our positivity arguments are shown with a gray cross (\textcolor{gray}{$\times$}).

\begin{landscape}

\begin{table}[t]																		
\begin{center}																		
\resizebox{1.3\textwidth}{!}{																		
\begin{tabular}{cccccccccccccccccc}																		
\toprule																		
 	&\boldmath{$\tilde{c}_{\phi^4D^4}^{(1)}$}	&\boldmath{$\tilde{c}_{\phi^4D^4}^{(2)}$}	&\boldmath{$\tilde{c}_{\phi^4D^4}^{(3)}$}	&\boldmath{$\tilde{c}_{l^2\phi^2D^3}^{(1)}$}	&\boldmath{$\tilde{c}_{l^2\phi^2D^3}^{(2)}$}	&\boldmath{$\tilde{c}_{l^2\phi^2D^3}^{(3)}$}	&\boldmath{$\tilde{c}_{l^2\phi^2D^3}^{(4)}$}	&\boldmath{$\tilde{c}_{e^2\phi^2D^3}^{(1)}$}	&\boldmath{$\tilde{c}_{e^2\phi^2D^3}^{(2)}$}	&\boldmath{$\tilde{c}_{q^2\phi^2D^3}^{(1)}$}	&\boldmath{$\tilde{c}_{q^2\phi^2D^3}^{(2)}$}	&\boldmath{$\tilde{c}_{q^2\phi^2D^3}^{(3)}$}	&\boldmath{$\tilde{c}_{q^2\phi^2D^3}^{(4)}$}	&\boldmath{$\tilde{c}_{u^2\phi^2D^3}^{(1)}$}	&\boldmath{$\tilde{c}_{u^2\phi^2D^3}^{(2)}$}	&\boldmath{$\tilde{c}_{d^2\phi^2D^3}^{(1)}$}	&\boldmath{$\tilde{c}_{d^2\phi^2D^3}^{(2)}$}	\\[0.2cm]
\boldmath{$\tilde{c}_{\phi^4D^4}^{(1)}$}	&\textcolor{gray}{$\times$}	&\textcolor{gray}{$\times$}	&\textcolor{gray}{$\times$}	&\textcolor{gray}{$\times$}	&\textcolor{gray}{$\times$}	&\textcolor{gray}{$\times$}	&\textcolor{gray}{$\times$}	&\textcolor{gray}{$\times$}	&\textcolor{gray}{$\times$}	&\textcolor{gray}{$\times$}	&\textcolor{gray}{$\times$}	&\textcolor{gray}{$\times$}	&\textcolor{gray}{$\times$}	&\textcolor{gray}{$\times$}	&\textcolor{gray}{$\times$}	&\textcolor{gray}{$\times$}	&\textcolor{gray}{$\times$}	\\[0.2cm]
\boldmath{$\tilde{c}_{\phi^4D^4}^{(2)}$}	&\textcolor{gray}{$\times$}	&\textcolor{gray}{$\times$}	&\textcolor{gray}{$\times$}	&\textcolor{gray}{$\times$}	&\textcolor{gray}{$\times$}	&\textcolor{gray}{$\times$}	&\textcolor{gray}{$\times$}	&\textcolor{gray}{$\times$}	&\textcolor{gray}{$\times$}	&\textcolor{gray}{$\times$}	&\textcolor{gray}{$\times$}	&\textcolor{gray}{$\times$}	&\textcolor{gray}{$\times$}	&\textcolor{gray}{$\times$}	&\textcolor{gray}{$\times$}	&\textcolor{gray}{$\times$}	&\textcolor{gray}{$\times$}	\\[0.2cm]
\boldmath{$\tilde{c}_{\phi^4D^4}^{(3)}$}	&\textcolor{gray}{$\times$}	&\textcolor{gray}{$\times$}	&\textcolor{gray}{$\times$}	&\textcolor{gray}{$\times$}	&\textcolor{gray}{$\times$}	&\textcolor{gray}{$\times$}	&\textcolor{gray}{$\times$}	&\textcolor{gray}{$\times$}	&\textcolor{gray}{$\times$}	&\textcolor{gray}{$\times$}	&\textcolor{gray}{$\times$}	&\textcolor{gray}{$\times$}	&\textcolor{gray}{$\times$}	&\textcolor{gray}{$\times$}	&\textcolor{gray}{$\times$}	&\textcolor{gray}{$\times$}	&\textcolor{gray}{$\times$}	\\[0.2cm]
\boldmath{$\tilde{c}_{G^2\phi^2D^2}^{(1)}$}	&\textcolor{blue}{$\boldsymbol{-}$}	&\textcolor{blue}{$\boldsymbol{-}$}	&\textcolor{blue}{$\boldsymbol{-}$}	&\textcolor{blue}{$\boldsymbol{-}$}	&\textcolor{blue}{$\boldsymbol{-}$}	&\fcolorbox{black}{LightCyan}{$\,0\,$}	&\fcolorbox{black}{LightCyan}{$\,0\,$}	&\textcolor{blue}{$\boldsymbol{-}$}	&\fcolorbox{black}{LightCyan}{$\,0\,$}	&\textcolor{blue}{$\boldsymbol{-}$}	&\textcolor{blue}{$\boldsymbol{-}$}	&\fcolorbox{black}{LightCyan}{$\,0\,$}	&\fcolorbox{black}{LightCyan}{$\,0\,$}	&\textcolor{blue}{$\boldsymbol{-}$}	&\fcolorbox{black}{LightCyan}{$\,0\,$}	&\textcolor{blue}{$\boldsymbol{-}$}	&\fcolorbox{black}{LightCyan}{$\,0\,$}	\\[0.2cm]
\boldmath{$\tilde{c}_{W^2\phi^2D^2}^{(1)}$}	&\textcolor{blue}{$\boldsymbol{-}$}	&\textcolor{blue}{$\boldsymbol{-}$}	&\textcolor{blue}{$\boldsymbol{-}$}	&\textcolor{blue}{$\boldsymbol{-}$}	&\textcolor{blue}{$\boldsymbol{-}$}	&\fcolorbox{black}{LightCyan}{$\,0\,$}	&\fcolorbox{black}{LightCyan}{$\,0\,$}	&\textcolor{blue}{$\boldsymbol{-}$}	&\fcolorbox{black}{LightCyan}{$\,0\,$}	&\textcolor{blue}{$\boldsymbol{-}$}	&\textcolor{blue}{$\boldsymbol{-}$}	&\fcolorbox{black}{LightCyan}{$\,0\,$}	&\fcolorbox{black}{LightCyan}{$\,0\,$}	&\textcolor{blue}{$\boldsymbol{-}$}	&\fcolorbox{black}{LightCyan}{$\,0\,$}	&\textcolor{blue}{$\boldsymbol{-}$}	&\fcolorbox{black}{LightCyan}{$\,0\,$}	\\[0.2cm]
\boldmath{$\tilde{c}_{B^2\phi^2D^2}^{(1)}$}	&\textcolor{blue}{$\boldsymbol{-}$}	&\textcolor{blue}{$\boldsymbol{-}$}	&\textcolor{blue}{$\boldsymbol{-}$}	&\textcolor{blue}{$\boldsymbol{-}$}	&\textcolor{blue}{$\boldsymbol{-}$}	&\fcolorbox{black}{LightCyan}{$\,0\,$}	&\fcolorbox{black}{LightCyan}{$\,0\,$}	&\textcolor{blue}{$\boldsymbol{-}$}	&\fcolorbox{black}{LightCyan}{$\,0\,$}	&\textcolor{blue}{$\boldsymbol{-}$}	&\textcolor{blue}{$\boldsymbol{-}$}	&\fcolorbox{black}{LightCyan}{$\,0\,$}	&\fcolorbox{black}{LightCyan}{$\,0\,$}	&\textcolor{blue}{$\boldsymbol{-}$}	&\fcolorbox{black}{LightCyan}{$\,0\,$}	&\textcolor{blue}{$\boldsymbol{-}$}	&\fcolorbox{black}{LightCyan}{$\,0\,$}	\\[0.2cm]
\boldmath{$\tilde{c}_{l^2\phi^2D^3}^{(1)}$}	&\textcolor{blue}{$\boldsymbol{-}$}	&\textcolor{blue}{$\boldsymbol{-}$}	&\textcolor{blue}{$\boldsymbol{-}$}	&\textcolor{gray}{$\times$}	&\textcolor{gray}{$\times$}	&\textcolor{gray}{$\times$}	&\textcolor{gray}{$\times$}	&\textcolor{blue}{$\boldsymbol{-}$}	&\fcolorbox{black}{LightCyan}{$\,0\,$}	&\textcolor{blue}{$\boldsymbol{-}$}	&\textcolor{blue}{$\boldsymbol{-}$}	&\fcolorbox{black}{LightCyan}{$\,0\,$}	&\fcolorbox{black}{LightCyan}{$\,0\,$}	&\textcolor{blue}{$\boldsymbol{-}$}	&\fcolorbox{black}{LightCyan}{$\,0\,$}	&\textcolor{blue}{$\boldsymbol{-}$}	&\fcolorbox{black}{LightCyan}{$\,0\,$}	\\[0.2cm]
\boldmath{$\tilde{c}_{l^2\phi^2D^3}^{(2)}$}	&\textcolor{blue}{$\boldsymbol{-}$}	&\textcolor{blue}{$\boldsymbol{-}$}	&\textcolor{blue}{$\boldsymbol{-}$}	&\textcolor{gray}{$\times$}	&\textcolor{gray}{$\times$}	&\textcolor{gray}{$\times$}	&\textcolor{gray}{$\times$}	&\textcolor{blue}{$\boldsymbol{-}$}	&\fcolorbox{black}{LightCyan}{$\,0\,$}	&\textcolor{blue}{$\boldsymbol{-}$}	&\textcolor{blue}{$\boldsymbol{-}$}	&\fcolorbox{black}{LightCyan}{$\,0\,$}	&\fcolorbox{black}{LightCyan}{$\,0\,$}	&\textcolor{blue}{$\boldsymbol{-}$}	&\fcolorbox{black}{LightCyan}{$\,0\,$}	&\textcolor{blue}{$\boldsymbol{-}$}	&\fcolorbox{black}{LightCyan}{$\,0\,$}	\\[0.2cm]
\boldmath{$\tilde{c}_{e^2\phi^2D^3}^{(1)}$}	&\textcolor{blue}{$\boldsymbol{-}$}	&\textcolor{blue}{$\boldsymbol{-}$}	&\textcolor{blue}{$\boldsymbol{-}$}	&\textcolor{blue}{$\boldsymbol{-}$}	&\textcolor{blue}{$\boldsymbol{-}$}	&\fcolorbox{black}{LightCyan}{$\,0\,$}	&\fcolorbox{black}{LightCyan}{$\,0\,$}	&\textcolor{gray}{$\times$}	&\textcolor{gray}{$\times$}	&\textcolor{blue}{$\boldsymbol{-}$}	&\textcolor{blue}{$\boldsymbol{-}$}	&\fcolorbox{black}{LightCyan}{$\,0\,$}	&\fcolorbox{black}{LightCyan}{$\,0\,$}	&\textcolor{blue}{$\boldsymbol{-}$}	&\fcolorbox{black}{LightCyan}{$\,0\,$}	&\textcolor{blue}{$\boldsymbol{-}$}	&\fcolorbox{black}{LightCyan}{$\,0\,$}	\\[0.2cm]
\boldmath{$\tilde{c}_{q^2\phi^2D^3}^{(1)}$}	&\textcolor{blue}{$\boldsymbol{-}$}	&\textcolor{blue}{$\boldsymbol{-}$}	&\textcolor{blue}{$\boldsymbol{-}$}	&\textcolor{blue}{$\boldsymbol{-}$}	&\textcolor{blue}{$\boldsymbol{-}$}	&\fcolorbox{black}{LightCyan}{$\,0\,$}	&\fcolorbox{black}{LightCyan}{$\,0\,$}	&\textcolor{blue}{$\boldsymbol{-}$}	&\fcolorbox{black}{LightCyan}{$\,0\,$}	&\textcolor{gray}{$\times$}	&\textcolor{gray}{$\times$}	&\textcolor{gray}{$\times$}	&\textcolor{gray}{$\times$}	&\textcolor{blue}{$\boldsymbol{-}$}	&\fcolorbox{black}{LightCyan}{$\,0\,$}	&\textcolor{blue}{$\boldsymbol{-}$}	&\fcolorbox{black}{LightCyan}{$\,0\,$}	\\[0.2cm]
\boldmath{$\tilde{c}_{q^2\phi^2D^3}^{(2)}$}	&\textcolor{blue}{$\boldsymbol{-}$}	&\textcolor{blue}{$\boldsymbol{-}$}	&\textcolor{blue}{$\boldsymbol{-}$}	&\textcolor{blue}{$\boldsymbol{-}$}	&\textcolor{blue}{$\boldsymbol{-}$}	&\fcolorbox{black}{LightCyan}{$\,0\,$}	&\fcolorbox{black}{LightCyan}{$\,0\,$}	&\textcolor{blue}{$\boldsymbol{-}$}	&\fcolorbox{black}{LightCyan}{$\,0\,$}	&\textcolor{gray}{$\times$}	&\textcolor{gray}{$\times$}	&\textcolor{gray}{$\times$}	&\textcolor{gray}{$\times$}	&\textcolor{blue}{$\boldsymbol{-}$}	&\fcolorbox{black}{LightCyan}{$\,0\,$}	&\textcolor{blue}{$\boldsymbol{-}$}	&\fcolorbox{black}{LightCyan}{$\,0\,$}	\\[0.2cm]
\boldmath{$\tilde{c}_{u^2\phi^2D^3}^{(1)}$}	&\textcolor{blue}{$\boldsymbol{-}$}	&\textcolor{blue}{$\boldsymbol{-}$}	&\textcolor{blue}{$\boldsymbol{-}$}	&\textcolor{blue}{$\boldsymbol{-}$}	&\textcolor{blue}{$\boldsymbol{-}$}	&\fcolorbox{black}{LightCyan}{$\,0\,$}	&\fcolorbox{black}{LightCyan}{$\,0\,$}	&\textcolor{blue}{$\boldsymbol{-}$}	&\fcolorbox{black}{LightCyan}{$\,0\,$}	&\textcolor{blue}{$\boldsymbol{-}$}	&\textcolor{blue}{$\boldsymbol{-}$}	&\fcolorbox{black}{LightCyan}{$\,0\,$}	&\fcolorbox{black}{LightCyan}{$\,0\,$}	&\textcolor{gray}{$\times$}	&\textcolor{gray}{$\times$}	&\textcolor{blue}{$\boldsymbol{-}$}	&\fcolorbox{black}{LightCyan}{$\,0\,$}	\\[0.2cm]
\boldmath{$\tilde{c}_{d^2\phi^2D^3}^{(1)}$}	&\textcolor{blue}{$\boldsymbol{-}$}	&\textcolor{blue}{$\boldsymbol{-}$}	&\textcolor{blue}{$\boldsymbol{-}$}	&\textcolor{blue}{$\boldsymbol{-}$}	&\textcolor{blue}{$\boldsymbol{-}$}	&\fcolorbox{black}{LightCyan}{$\,0\,$}	&\fcolorbox{black}{LightCyan}{$\,0\,$}	&\textcolor{blue}{$\boldsymbol{-}$}	&\fcolorbox{black}{LightCyan}{$\,0\,$}	&\textcolor{blue}{$\boldsymbol{-}$}	&\textcolor{blue}{$\boldsymbol{-}$}	&\fcolorbox{black}{LightCyan}{$\,0\,$}	&\fcolorbox{black}{LightCyan}{$\,0\,$}	&\textcolor{blue}{$\boldsymbol{-}$}	&\fcolorbox{black}{LightCyan}{$\,0\,$}	&\textcolor{gray}{$\times$}	&\textcolor{gray}{$\times$}	\\[0.2cm]
\boldmath{$\tilde{c}_{q^2G^2D}^{(1)}$}	&0	&0	&0	&0	&0	&0	&0	&0	&0	&\textcolor{blue}{$\boldsymbol{-}$}	&\textcolor{blue}{$\boldsymbol{-}$}	&\fcolorbox{black}{LightCyan}{$\,0\,$}	&\fcolorbox{black}{LightCyan}{$\,0\,$}	&0	&0	&0	&0	\\[0.2cm]
\boldmath{$\tilde{c}_{q^2G^2D}^{(2)}$}	&0	&0	&0	&0	&0	&0	&0	&0	&0	&\textcolor{blue}{$\boldsymbol{-}$}	&\textcolor{blue}{$\boldsymbol{-}$}	&\fcolorbox{black}{LightCyan}{$\,0\,$}	&\fcolorbox{black}{LightCyan}{$\,0\,$}	&0	&0	&0	&0	\\[0.2cm]
\boldmath{$\tilde{c}_{q^2W^2D}^{(1)}$}	&0	&0	&0	&0	&0	&0	&0	&0	&0	&\textcolor{blue}{$\boldsymbol{-}$}	&\textcolor{blue}{$\boldsymbol{-}$}	&\fcolorbox{black}{LightCyan}{$\,0\,$}	&\fcolorbox{black}{LightCyan}{$\,0\,$}	&0	&0	&0	&0	\\[0.2cm]
\boldmath{$\tilde{c}_{q^2B^2D}^{(1)}$}	&0	&0	&0	&0	&0	&0	&0	&0	&0	&\textcolor{blue}{$\boldsymbol{-}$}	&\textcolor{blue}{$\boldsymbol{-}$}	&\fcolorbox{black}{LightCyan}{$\,0\,$}	&\fcolorbox{black}{LightCyan}{$\,0\,$}	&0	&0	&0	&0	\\[0.2cm]
\boldmath{$\tilde{c}_{u^2G^2D}^{(1)}$}	&0	&0	&0	&0	&0	&0	&0	&0	&0	&0	&0	&0	&0	&\textcolor{blue}{$\boldsymbol{-}$}	&\fcolorbox{black}{LightCyan}{$\,0\,$}	&0	&0	\\[0.2cm]
\boldmath{$\tilde{c}_{u^2G^2D}^{(2)}$}	&0	&0	&0	&0	&0	&0	&0	&0	&0	&0	&0	&0	&0	&\textcolor{blue}{$\boldsymbol{-}$}	&\fcolorbox{black}{LightCyan}{$\,0\,$}	&0	&0	\\[0.2cm]
\boldmath{$\tilde{c}_{u^2W^2D}$}	&0	&0	&0	&0	&0	&0	&0	&0	&0	&0	&0	&0	&0	&\textcolor{blue}{$\boldsymbol{-}$}	&\fcolorbox{black}{LightCyan}{$\,0\,$}	&0	&0	\\[0.2cm]
\boldmath{$\tilde{c}_{u^2B^2D}$}	&0	&0	&0	&0	&0	&0	&0	&0	&0	&0	&0	&0	&0	&\textcolor{blue}{$\boldsymbol{-}$}	&\fcolorbox{black}{LightCyan}{$\,0\,$}	&0	&0	\\[0.2cm]
\boldmath{$\tilde{c}_{d^2G^2D}^{(1)}$}	&0	&0	&0	&0	&0	&0	&0	&0	&0	&0	&0	&0	&0	&0	&0	&\textcolor{blue}{$\boldsymbol{-}$}	&\fcolorbox{black}{LightCyan}{$\,0\,$}	\\[0.2cm]
\boldmath{$\tilde{c}_{d^2G^2D}^{(2)}$}	&0	&0	&0	&0	&0	&0	&0	&0	&0	&0	&0	&0	&0	&0	&0	&\textcolor{blue}{$\boldsymbol{-}$}	&\fcolorbox{black}{LightCyan}{$\,0\,$}	\\[0.2cm]
\boldmath{$\tilde{c}_{d^2W^2D}$}	&0	&0	&0	&0	&0	&0	&0	&0	&0	&0	&0	&0	&0	&0	&0	&\textcolor{blue}{$\boldsymbol{-}$}	&\fcolorbox{black}{LightCyan}{$\,0\,$}	\\[0.2cm]
\boldmath{$\tilde{c}_{d^2B^2D}$}	&0	&0	&0	&0	&0	&0	&0	&0	&0	&0	&0	&0	&0	&0	&0	&\textcolor{blue}{$\boldsymbol{-}$}	&\fcolorbox{black}{LightCyan}{$\,0\,$}	\\[0.2cm]
\boldmath{$\tilde{c}_{l^2G^2D}$}	&0	&0	&0	&\textcolor{blue}{$\boldsymbol{-}$}	&\textcolor{blue}{$\boldsymbol{-}$}	&\fcolorbox{black}{LightCyan}{$\,0\,$}	&\fcolorbox{black}{LightCyan}{$\,0\,$}	&0	&0	&0	&0	&0	&0	&0	&0	&0	&0	\\[0.2cm]
\boldmath{$\tilde{c}_{l^2W^2D}$}	&0	&0	&0	&\textcolor{blue}{$\boldsymbol{-}$}	&\textcolor{blue}{$\boldsymbol{-}$}	&\fcolorbox{black}{LightCyan}{$\,0\,$}	&\fcolorbox{black}{LightCyan}{$\,0\,$}	&0	&0	&0	&0	&0	&0	&0	&0	&0	&0	\\[0.2cm]
\boldmath{$\tilde{c}_{l^2B^2D}$}	&0	&0	&0	&\textcolor{blue}{$\boldsymbol{-}$}	&\textcolor{blue}{$\boldsymbol{-}$}	&\fcolorbox{black}{LightCyan}{$\,0\,$}	&\fcolorbox{black}{LightCyan}{$\,0\,$}	&0	&0	&0	&0	&0	&0	&0	&0	&0	&0	\\[0.2cm]
\boldmath{$\tilde{c}_{e^2G^2D}$}	&0	&0	&0	&0	&0	&0	&0	&\textcolor{blue}{$\boldsymbol{-}$}	&\fcolorbox{black}{LightCyan}{$\,0\,$}	&0	&0	&0	&0	&0	&0	&0	&0	\\[0.2cm]
\boldmath{$\tilde{c}_{e^2W^2D}$}	&0	&0	&0	&0	&0	&0	&0	&\textcolor{blue}{$\boldsymbol{-}$}	&\fcolorbox{black}{LightCyan}{$\,0\,$}	&0	&0	&0	&0	&0	&0	&0	&0	\\[0.2cm]
\boldmath{$\tilde{c}_{e^2B^2D}$}	&0	&0	&0	&0	&0	&0	&0	&\textcolor{blue}{$\boldsymbol{-}$}	&\fcolorbox{black}{LightCyan}{$\,0\,$}	&0	&0	&0	&0	&0	&0	&0	&0	\\[0.2cm]
\bottomrule																		
\end{tabular}}																		
\end{center}																		
\caption{\it Mixing between operators containing at most two fermions.}\label{tab:adm1}																		
\end{table}

\begin{table}[t]															
\begin{center}															
\resizebox{1.3\textwidth}{!}{															
\begin{tabular}{ccccccccccccccc}															
\toprule															
 	&\boldmath{$\tilde{c}_{l^2\phi^2D^3}^{(1)}$}	&\boldmath{$\tilde{c}_{l^2\phi^2D^3}^{(2)}$}	&\boldmath{$\tilde{c}_{l^2\phi^2D^3}^{(3)}$}	&\boldmath{$\tilde{c}_{l^2\phi^2D^3}^{(4)}$}	&\boldmath{$\tilde{c}_{e^2\phi^2D^3}^{(1)}$}	&\boldmath{$\tilde{c}_{e^2\phi^2D^3}^{(2)}$}	&\boldmath{$\tilde{c}_{q^2\phi^2D^3}^{(1)}$}	&\boldmath{$\tilde{c}_{q^2\phi^2D^3}^{(2)}$}	&\boldmath{$\tilde{c}_{q^2\phi^2D^3}^{(3)}$}	&\boldmath{$\tilde{c}_{q^2\phi^2D^3}^{(4)}$}	&\boldmath{$\tilde{c}_{u^2\phi^2D^3}^{(1)}$}	&\boldmath{$\tilde{c}_{u^2\phi^2D^3}^{(2)}$}	&\boldmath{$\tilde{c}_{d^2\phi^2D^3}^{(1)}$}	&\boldmath{$\tilde{c}_{d^2\phi^2D^3}^{(2)}$}	\\[0.2cm]
\boldmath{$\tilde{c}_{l^4D^2}^{(1)}$}	&\textcolor{gray}{$\times$}	&\textcolor{gray}{$\times$}	&\textcolor{gray}{$\times$}	&\textcolor{gray}{$\times$}	&0	&0	&0	&0	&0	&0	&0	&0	&0	&0	\\[0.2cm]
\boldmath{$\tilde{c}_{l^4D^2}^{(2)}$}	&\textcolor{gray}{$\times$}	&\textcolor{gray}{$\times$}	&\textcolor{gray}{$\times$}	&\textcolor{gray}{$\times$}	&0	&0	&0	&0	&0	&0	&0	&0	&0	&0	\\[0.2cm]
\boldmath{$\tilde{c}_{q^4D^2}^{(1)}$}	&0	&0	&0	&0	&0	&0	&\textcolor{gray}{$\times$}	&\textcolor{gray}{$\times$}	&\textcolor{gray}{$\times$}	&\textcolor{gray}{$\times$}	&0	&0	&0	&0	\\[0.2cm]
\boldmath{$\tilde{c}_{q^4D^2}^{(2)}$}	&0	&0	&0	&0	&0	&0	&\textcolor{gray}{$\times$}	&\textcolor{gray}{$\times$}	&\textcolor{gray}{$\times$}	&\textcolor{gray}{$\times$}	&0	&0	&0	&0	\\[0.2cm]
\boldmath{$\tilde{c}_{q^4D^2}^{(3)}$}	&0	&0	&0	&0	&0	&0	&\textcolor{gray}{$\times$}	&\textcolor{gray}{$\times$}	&\textcolor{gray}{$\times$}	&\textcolor{gray}{$\times$}	&0	&0	&0	&0	\\[0.2cm]
\boldmath{$\tilde{c}_{q^4D^2}^{(4)}$}	&0	&0	&0	&0	&0	&0	&\textcolor{gray}{$\times$}	&\textcolor{gray}{$\times$}	&\textcolor{gray}{$\times$}	&\textcolor{gray}{$\times$}	&0	&0	&0	&0	\\[0.2cm]
\boldmath{$\tilde{c}_{l^2q^2D^2}^{(1)}$}	&\textcolor{blue}{$\boldsymbol{-}$}	&\textcolor{blue}{$\boldsymbol{-}$}	&\fcolorbox{black}{LightCyan}{$\,0\,$}	&\fcolorbox{black}{LightCyan}{$\,0\,$}	&0	&0	&\textcolor{blue}{$\boldsymbol{-}$}	&\textcolor{blue}{$\boldsymbol{-}$}	&\fcolorbox{black}{LightCyan}{$\,0\,$}	&\fcolorbox{black}{LightCyan}{$\,0\,$}	&0	&0	&0	&0	\\[0.2cm]
\boldmath{$\tilde{c}_{l^2q^2D^2}^{(2)}$}	&\textcolor{blue}{$\boldsymbol{-}$}	&\textcolor{blue}{$\boldsymbol{-}$}	&\fcolorbox{black}{LightCyan}{$\,0\,$}	&\fcolorbox{black}{LightCyan}{$\,0\,$}	&0	&0	&\textcolor{blue}{$\boldsymbol{-}$}	&\textcolor{blue}{$\boldsymbol{-}$}	&\fcolorbox{black}{LightCyan}{$\,0\,$}	&\fcolorbox{black}{LightCyan}{$\,0\,$}	&0	&0	&0	&0	\\[0.2cm]
\boldmath{$\tilde{c}_{e^4D^2}$}	&0	&0	&0	&0	&\textcolor{gray}{$\times$}	&\textcolor{gray}{$\times$}	&0	&0	&0	&0	&0	&0	&0	&0	\\[0.2cm]
\boldmath{$\tilde{c}_{u^4D^2}^{(1)}$}	&0	&0	&0	&0	&0	&0	&0	&0	&0	&0	&\textcolor{gray}{$\times$}	&\textcolor{gray}{$\times$}	&0	&0	\\[0.2cm]
\boldmath{$\tilde{c}_{u^4D^2}^{(2)}$}	&0	&0	&0	&0	&0	&0	&0	&0	&0	&0	&\textcolor{gray}{$\times$}	&\textcolor{gray}{$\times$}	&0	&0	\\[0.2cm]
\boldmath{$\tilde{c}_{d^4D^2}^{(1)}$}	&0	&0	&0	&0	&0	&0	&0	&0	&0	&0	&0	&0	&\textcolor{gray}{$\times$}	&\textcolor{gray}{$\times$}	\\[0.2cm]
\boldmath{$\tilde{c}_{d^4D^2}^{(2)}$}	&0	&0	&0	&0	&0	&0	&0	&0	&0	&0	&0	&0	&\textcolor{gray}{$\times$}	&\textcolor{gray}{$\times$}	\\[0.2cm]
\boldmath{$\tilde{c}_{e^2u^2D^2}^{(1)}$}	&0	&0	&0	&0	&\textcolor{blue}{$\boldsymbol{-}$}	&\fcolorbox{black}{LightCyan}{$\,0\,$}	&0	&0	&0	&0	&\textcolor{blue}{$\boldsymbol{-}$}	&\fcolorbox{black}{LightCyan}{$\,0\,$}	&0	&0	\\[0.2cm]
\boldmath{$\tilde{c}_{e^2d^2D^2}^{(1)}$}	&0	&0	&0	&0	&\textcolor{blue}{$\boldsymbol{-}$}	&\fcolorbox{black}{LightCyan}{$\,0\,$}	&0	&0	&0	&0	&0	&0	&\textcolor{blue}{$\boldsymbol{-}$}	&\fcolorbox{black}{LightCyan}{$\,0\,$}	\\[0.2cm]
\boldmath{$\tilde{c}_{u^2d^2D^2}^{(1)}$}	&0	&0	&0	&0	&0	&0	&0	&0	&0	&0	&\textcolor{blue}{$\boldsymbol{-}$}	&\fcolorbox{black}{LightCyan}{$\,0\,$}	&\textcolor{blue}{$\boldsymbol{-}$}	&\fcolorbox{black}{LightCyan}{$\,0\,$}	\\[0.2cm]
\boldmath{$\tilde{c}_{u^2d^2D^2}^{(2)}$}	&0	&0	&0	&0	&0	&0	&0	&0	&0	&0	&\textcolor{blue}{$\boldsymbol{-}$}	&\fcolorbox{black}{LightCyan}{$\,0\,$}	&\textcolor{blue}{$\boldsymbol{-}$}	&\fcolorbox{black}{LightCyan}{$\,0\,$}	\\[0.2cm]
\boldmath{$\tilde{c}_{l^2e^2D^2}^{(1)}$}	&\textcolor{blue}{$\boldsymbol{-}$}	&\textcolor{blue}{$\boldsymbol{-}$}	&\fcolorbox{black}{LightCyan}{$\,0\,$}	&\fcolorbox{black}{LightCyan}{$\,0\,$}	&\textcolor{blue}{$\boldsymbol{-}$}	&\fcolorbox{black}{LightCyan}{$\,0\,$}	&0	&0	&0	&0	&0	&0	&0	&0	\\[0.2cm]
\boldmath{$\tilde{c}_{l^2u^2D^2}^{(1)}$}	&\textcolor{blue}{$\boldsymbol{-}$}	&\textcolor{blue}{$\boldsymbol{-}$}	&\fcolorbox{black}{LightCyan}{$\,0\,$}	&\fcolorbox{black}{LightCyan}{$\,0\,$}	&0	&0	&0	&0	&0	&0	&\textcolor{blue}{$\boldsymbol{-}$}	&\fcolorbox{black}{LightCyan}{$\,0\,$}	&0	&0	\\[0.2cm]
\boldmath{$\tilde{c}_{l^2d^2D^2}^{(1)}$}	&\textcolor{blue}{$\boldsymbol{-}$}	&\textcolor{blue}{$\boldsymbol{-}$}	&\fcolorbox{black}{LightCyan}{$\,0\,$}	&\fcolorbox{black}{LightCyan}{$\,0\,$}	&0	&0	&0	&0	&0	&0	&0	&0	&\textcolor{blue}{$\boldsymbol{-}$}	&\fcolorbox{black}{LightCyan}{$\,0\,$}	\\[0.2cm]
\boldmath{$\tilde{c}_{q^2e^2D^2}^{(1)}$}	&0	&0	&0	&0	&\textcolor{blue}{$\boldsymbol{-}$}	&\fcolorbox{black}{LightCyan}{$\,0\,$}	&\textcolor{blue}{$\boldsymbol{-}$}	&\textcolor{blue}{$\boldsymbol{-}$}	&\fcolorbox{black}{LightCyan}{$\,0\,$}	&\fcolorbox{black}{LightCyan}{$\,0\,$}	&0	&0	&0	&0	\\[0.2cm]
\boldmath{$\tilde{c}_{q^2u^2D^2}^{(1)}$}	&0	&0	&0	&0	&0	&0	&\textcolor{blue}{$\boldsymbol{-}$}	&\textcolor{blue}{$\boldsymbol{-}$}	&\fcolorbox{black}{LightCyan}{$\,0\,$}	&\fcolorbox{black}{LightCyan}{$\,0\,$}	&\textcolor{blue}{$\boldsymbol{-}$}	&\fcolorbox{black}{LightCyan}{$\,0\,$}	&0	&0	\\[0.2cm]
\boldmath{$\tilde{c}_{q^2u^2D^2}^{(2)}$}	&0	&0	&0	&0	&0	&0	&\textcolor{blue}{$\boldsymbol{-}$}	&\textcolor{blue}{$\boldsymbol{-}$}	&\fcolorbox{black}{LightCyan}{$\,0\,$}	&\fcolorbox{black}{LightCyan}{$\,0\,$}	&\textcolor{blue}{$\boldsymbol{-}$}	&\fcolorbox{black}{LightCyan}{$\,0\,$}	&0	&0	\\[0.2cm]
\boldmath{$\tilde{c}_{q^2d^2D^2}^{(1)}$}	&0	&0	&0	&0	&0	&0	&\textcolor{blue}{$\boldsymbol{-}$}	&\textcolor{blue}{$\boldsymbol{-}$}	&\fcolorbox{black}{LightCyan}{$\,0\,$}	&\fcolorbox{black}{LightCyan}{$\,0\,$}	&0	&0	&\textcolor{blue}{$\boldsymbol{-}$}	&\fcolorbox{black}{LightCyan}{$\,0\,$}	\\[0.2cm]
\boldmath{$\tilde{c}_{q^2d^2D^2}^{(2)}$}	&0	&0	&0	&0	&0	&0	&\textcolor{blue}{$\boldsymbol{-}$}	&\textcolor{blue}{$\boldsymbol{-}$}	&\fcolorbox{black}{LightCyan}{$\,0\,$}	&\fcolorbox{black}{LightCyan}{$\,0\,$}	&0	&0	&\textcolor{blue}{$\boldsymbol{-}$}	&\fcolorbox{black}{LightCyan}{$\,0\,$}	\\[0.2cm]
\bottomrule															
\end{tabular}}															
\end{center}															
\caption{\it Mixing of two fermion operators into four-fermion ones.}\label{tab:adm3}															
\end{table}

\begin{table}[t]																																								
\begin{center}																																								
\resizebox{1.3\textwidth}{!}{																																								
\begin{tabular}{cccccccccccccccccccccccccccccccccccccccc}																																								
\toprule																																								
 	&\boldmath{$\tilde{c}_{l^4D^2}^{(1)}$}	&\boldmath{$\tilde{c}_{l^4D^2}^{(2)}$}	&\boldmath{$\tilde{c}_{q^4D^2}^{(1)}$}	&\boldmath{$\tilde{c}_{q^4D^2}^{(2)}$}	&\boldmath{$\tilde{c}_{q^4D^2}^{(3)}$}	&\boldmath{$\tilde{c}_{q^4D^2}^{(4)}$}	&\boldmath{$\tilde{c}_{l^2q^2D^2}^{(1)}$}	&\boldmath{$\tilde{c}_{l^2q^2D^2}^{(2)}$}	&\boldmath{$\tilde{c}_{l^2q^2D^2}^{(3)}$}	&\boldmath{$\tilde{c}_{l^2q^2D^2}^{(4)}$}	&\boldmath{$\tilde{c}_{e^4D^2}$}	&\boldmath{$\tilde{c}_{u^4D^2}^{(1)}$}	&\boldmath{$\tilde{c}_{u^4D^2}^{(2)}$}	&\boldmath{$\tilde{c}_{d^4D^2}^{(1)}$}	&\boldmath{$\tilde{c}_{d^4D^2}^{(2)}$}	&\boldmath{$\tilde{c}_{e^2u^2D^2}^{(1)}$}	&\boldmath{$\tilde{c}_{e^2u^2D^2}^{(2)}$}	&\boldmath{$\tilde{c}_{e^2d^2D^2}^{(1)}$}	&\boldmath{$\tilde{c}_{e^2d^2D^2}^{(2)}$}	&\boldmath{$\tilde{c}_{u^2d^2D^2}^{(1)}$}	&\boldmath{$\tilde{c}_{u^2d^2D^2}^{(2)}$}	&\boldmath{$\tilde{c}_{u^2d^2D^2}^{(3)}$}	&\boldmath{$\tilde{c}_{u^2d^2D^2}^{(4)}$}	&\boldmath{$\tilde{c}_{l^2e^2D^2}^{(1)}$}	&\boldmath{$\tilde{c}_{l^2e^2D^2}^{(2)}$}	&\boldmath{$\tilde{c}_{l^2u^2D^2}^{(1)}$}	&\boldmath{$\tilde{c}_{l^2u^2D^2}^{(2)}$}	&\boldmath{$\tilde{c}_{l^2d^2D^2}^{(1)}$}	&\boldmath{$\tilde{c}_{l^2d^2D^2}^{(2)}$}	&\boldmath{$\tilde{c}_{q^2e^2D^2}^{(1)}$}	&\boldmath{$\tilde{c}_{q^2e^2D^2}^{(2)}$}	&\boldmath{$\tilde{c}_{q^2u^2D^2}^{(1)}$}	&\boldmath{$\tilde{c}_{q^2u^2D^2}^{(2)}$}	&\boldmath{$\tilde{c}_{q^2u^2D^2}^{(3)}$}	&\boldmath{$\tilde{c}_{q^2u^2D^2}^{(4)}$}	&\boldmath{$\tilde{c}_{q^2d^2D^2}^{(1)}$}	&\boldmath{$\tilde{c}_{q^2d^2D^2}^{(2)}$}	&\boldmath{$\tilde{c}_{q^2d^2D^2}^{(3)}$}	&\boldmath{$\tilde{c}_{q^2d^2D^2}^{(4)}$}	\\[0.2cm]
\boldmath{$\tilde{c}_{l^2\phi^2D^3}^{(1)}$}	&\textcolor{blue}{$\boldsymbol{-}$}	&\textcolor{blue}{$\boldsymbol{-}$}	&0	&0	&0	&0	&\textcolor{gray}{$\times$}	&\textcolor{gray}{$\times$}	&\textcolor{gray}{$\times$}	&\textcolor{gray}{$\times$}	&0	&0	&0	&0	&0	&0	&0	&0	&0	&0	&0	&0	&0	&\textcolor{gray}{$\times$}	&\textcolor{gray}{$\times$}	&\textcolor{gray}{$\times$}	&\textcolor{gray}{$\times$}	&\textcolor{gray}{$\times$}	&\textcolor{gray}{$\times$}	&0	&0	&0	&0	&0	&0	&0	&0	&0	&0	\\[0.2cm]
\boldmath{$\tilde{c}_{l^2\phi^2D^3}^{(2)}$}	&\textcolor{blue}{$\boldsymbol{-}$}	&\textcolor{blue}{$\boldsymbol{-}$}	&0	&0	&0	&0	&\textcolor{gray}{$\times$}	&\textcolor{gray}{$\times$}	&\textcolor{gray}{$\times$}	&\textcolor{gray}{$\times$}	&0	&0	&0	&0	&0	&0	&0	&0	&0	&0	&0	&0	&0	&\textcolor{gray}{$\times$}	&\textcolor{gray}{$\times$}	&\textcolor{gray}{$\times$}	&\textcolor{gray}{$\times$}	&\textcolor{gray}{$\times$}	&\textcolor{gray}{$\times$}	&0	&0	&0	&0	&0	&0	&0	&0	&0	&0	\\[0.2cm]
\boldmath{$\tilde{c}_{e^2\phi^2D^3}^{(1)}$}	&0	&0	&0	&0	&0	&0	&0	&0	&0	&0	&\textcolor{blue}{$\boldsymbol{-}$}	&0	&0	&0	&0	&\textcolor{blue}{$\boldsymbol{-}$}	&\textcolor{gray}{$\times$}	&\textcolor{blue}{$\boldsymbol{-}$}	&\textcolor{gray}{$\times$}	&0	&0	&0	&0	&\textcolor{gray}{$\times$}	&\textcolor{gray}{$\times$}	&0	&0	&0	&0	&\textcolor{gray}{$\times$}	&\textcolor{gray}{$\times$}	&0	&0	&0	&0	&0	&0	&0	&0	\\[0.2cm]
\boldmath{$\tilde{c}_{q^2\phi^2D^3}^{(1)}$}	&0	&0	&\textcolor{blue}{$\boldsymbol{-}$}	&\textcolor{blue}{$\boldsymbol{-}$}	&\textcolor{blue}{$\boldsymbol{-}$}	&\textcolor{blue}{$\boldsymbol{-}$}	&\textcolor{gray}{$\times$}	&\textcolor{gray}{$\times$}	&\textcolor{gray}{$\times$}	&\textcolor{gray}{$\times$}	&0	&0	&0	&0	&0	&0	&0	&0	&0	&0	&0	&0	&0	&0	&0	&0	&0	&0	&0	&\textcolor{gray}{$\times$}	&\textcolor{gray}{$\times$}	&\textcolor{gray}{$\times$}	&\textcolor{gray}{$\times$}	&\textcolor{gray}{$\times$}	&\textcolor{gray}{$\times$}	&\textcolor{gray}{$\times$}	&\textcolor{gray}{$\times$}	&\textcolor{gray}{$\times$}	&\textcolor{gray}{$\times$}	\\[0.2cm]
\boldmath{$\tilde{c}_{q^2\phi^2D^3}^{(2)}$}	&0	&0	&\textcolor{blue}{$\boldsymbol{-}$}	&\textcolor{blue}{$\boldsymbol{-}$}	&\textcolor{blue}{$\boldsymbol{-}$}	&\textcolor{blue}{$\boldsymbol{-}$}	&\textcolor{gray}{$\times$}	&\textcolor{gray}{$\times$}	&\textcolor{gray}{$\times$}	&\textcolor{gray}{$\times$}	&0	&0	&0	&0	&0	&0	&0	&0	&0	&0	&0	&0	&0	&0	&0	&0	&0	&0	&0	&\textcolor{gray}{$\times$}	&\textcolor{gray}{$\times$}	&\textcolor{gray}{$\times$}	&\textcolor{gray}{$\times$}	&\textcolor{gray}{$\times$}	&\textcolor{gray}{$\times$}	&\textcolor{gray}{$\times$}	&\textcolor{gray}{$\times$}	&\textcolor{gray}{$\times$}	&\textcolor{gray}{$\times$}	\\[0.2cm]
\boldmath{$\tilde{c}_{u^2\phi^2D^3}^{(1)}$}	&0	&0	&0	&0	&0	&0	&0	&0	&0	&0	&0	&\textcolor{blue}{$\boldsymbol{-}$}	&\textcolor{blue}{$\boldsymbol{-}$}	&0	&0	&\textcolor{blue}{$\boldsymbol{-}$}	&\textcolor{gray}{$\times$}	&0	&0	&\textcolor{gray}{$\times$}	&\textcolor{gray}{$\times$}	&\textcolor{gray}{$\times$}	&\textcolor{gray}{$\times$}	&0	&0	&\textcolor{gray}{$\times$}	&\textcolor{gray}{$\times$}	&0	&0	&0	&0	&\textcolor{gray}{$\times$}	&\textcolor{gray}{$\times$}	&\textcolor{gray}{$\times$}	&\textcolor{gray}{$\times$}	&0	&0	&0	&0	\\[0.2cm]
\boldmath{$\tilde{c}_{d^2\phi^2D^3}^{(1)}$}	&0	&0	&0	&0	&0	&0	&0	&0	&0	&0	&0	&0	&0	&\textcolor{blue}{$\boldsymbol{-}$}	&\textcolor{blue}{$\boldsymbol{-}$}	&0	&0	&\textcolor{blue}{$\boldsymbol{-}$}	&\textcolor{gray}{$\times$}	&\textcolor{gray}{$\times$}	&\textcolor{gray}{$\times$}	&\textcolor{gray}{$\times$}	&\textcolor{gray}{$\times$}	&0	&0	&0	&0	&\textcolor{gray}{$\times$}	&\textcolor{gray}{$\times$}	&0	&0	&0	&0	&0	&0	&\textcolor{gray}{$\times$}	&\textcolor{gray}{$\times$}	&\textcolor{gray}{$\times$}	&\textcolor{gray}{$\times$}	\\[0.2cm]
\boldmath{$\tilde{c}_{q^2G^2D}^{(1)}$}	&0	&0	&\textcolor{blue}{$\boldsymbol{-}$}	&\textcolor{blue}{$\boldsymbol{-}$}	&\textcolor{blue}{$\boldsymbol{-}$}	&\textcolor{blue}{$\boldsymbol{-}$}	&\textcolor{gray}{$\times$}	&\textcolor{gray}{$\times$}	&\textcolor{gray}{$\times$}	&\textcolor{gray}{$\times$}	&0	&0	&0	&0	&0	&0	&0	&0	&0	&0	&0	&0	&0	&0	&0	&0	&0	&0	&0	&\textcolor{gray}{$\times$}	&\textcolor{gray}{$\times$}	&\textcolor{gray}{$\times$}	&\textcolor{gray}{$\times$}	&\textcolor{gray}{$\times$}	&\textcolor{gray}{$\times$}	&\textcolor{gray}{$\times$}	&\textcolor{gray}{$\times$}	&\textcolor{gray}{$\times$}	&\textcolor{gray}{$\times$}	\\[0.2cm]
\boldmath{$\tilde{c}_{q^2G^2D}^{(2)}$}	&0	&0	&\textcolor{blue}{$\boldsymbol{-}$}	&\textcolor{blue}{$\boldsymbol{-}$}	&\textcolor{blue}{$\boldsymbol{-}$}	&\textcolor{blue}{$\boldsymbol{-}$}	&\textcolor{gray}{$\times$}	&\textcolor{gray}{$\times$}	&\textcolor{gray}{$\times$}	&\textcolor{gray}{$\times$}	&0	&0	&0	&0	&0	&0	&0	&0	&0	&0	&0	&0	&0	&0	&0	&0	&0	&0	&0	&\textcolor{gray}{$\times$}	&\textcolor{gray}{$\times$}	&\textcolor{gray}{$\times$}	&\textcolor{gray}{$\times$}	&\textcolor{gray}{$\times$}	&\textcolor{gray}{$\times$}	&\textcolor{gray}{$\times$}	&\textcolor{gray}{$\times$}	&\textcolor{gray}{$\times$}	&\textcolor{gray}{$\times$}	\\[0.2cm]
\boldmath{$\tilde{c}_{q^2W^2D}^{(1)}$}	&0	&0	&\textcolor{blue}{$\boldsymbol{-}$}	&\textcolor{blue}{$\boldsymbol{-}$}	&\textcolor{blue}{$\boldsymbol{-}$}	&\textcolor{blue}{$\boldsymbol{-}$}	&\textcolor{gray}{$\times$}	&\textcolor{gray}{$\times$}	&\textcolor{gray}{$\times$}	&\textcolor{gray}{$\times$}	&0	&0	&0	&0	&0	&0	&0	&0	&0	&0	&0	&0	&0	&0	&0	&0	&0	&0	&0	&\textcolor{gray}{$\times$}	&\textcolor{gray}{$\times$}	&\textcolor{gray}{$\times$}	&\textcolor{gray}{$\times$}	&\textcolor{gray}{$\times$}	&\textcolor{gray}{$\times$}	&\textcolor{gray}{$\times$}	&\textcolor{gray}{$\times$}	&\textcolor{gray}{$\times$}	&\textcolor{gray}{$\times$}	\\[0.2cm]
\boldmath{$\tilde{c}_{q^2B^2D}^{(1)}$}	&0	&0	&\textcolor{blue}{$\boldsymbol{-}$}	&\textcolor{blue}{$\boldsymbol{-}$}	&\textcolor{blue}{$\boldsymbol{-}$}	&\textcolor{blue}{$\boldsymbol{-}$}	&\textcolor{gray}{$\times$}	&\textcolor{gray}{$\times$}	&\textcolor{gray}{$\times$}	&\textcolor{gray}{$\times$}	&0	&0	&0	&0	&0	&0	&0	&0	&0	&0	&0	&0	&0	&0	&0	&0	&0	&0	&0	&\textcolor{gray}{$\times$}	&\textcolor{gray}{$\times$}	&\textcolor{gray}{$\times$}	&\textcolor{gray}{$\times$}	&\textcolor{gray}{$\times$}	&\textcolor{gray}{$\times$}	&\textcolor{gray}{$\times$}	&\textcolor{gray}{$\times$}	&\textcolor{gray}{$\times$}	&\textcolor{gray}{$\times$}	\\[0.2cm]
\boldmath{$\tilde{c}_{u^2G^2D}^{(1)}$}	&0	&0	&0	&0	&0	&0	&0	&0	&0	&0	&0	&\textcolor{blue}{$\boldsymbol{-}$}	&\textcolor{blue}{$\boldsymbol{-}$}	&0	&0	&\textcolor{blue}{$\boldsymbol{-}$}	&\textcolor{gray}{$\times$}	&0	&0	&\textcolor{gray}{$\times$}	&\textcolor{gray}{$\times$}	&\textcolor{gray}{$\times$}	&\textcolor{gray}{$\times$}	&0	&0	&\textcolor{gray}{$\times$}	&\textcolor{gray}{$\times$}	&0	&0	&0	&0	&\textcolor{gray}{$\times$}	&\textcolor{gray}{$\times$}	&\textcolor{gray}{$\times$}	&\textcolor{gray}{$\times$}	&0	&0	&0	&0	\\[0.2cm]
\boldmath{$\tilde{c}_{u^2G^2D}^{(2)}$}	&0	&0	&0	&0	&0	&0	&0	&0	&0	&0	&0	&\textcolor{blue}{$\boldsymbol{-}$}	&\textcolor{blue}{$\boldsymbol{-}$}	&0	&0	&\textcolor{blue}{$\boldsymbol{-}$}	&\textcolor{gray}{$\times$}	&0	&0	&\textcolor{gray}{$\times$}	&\textcolor{gray}{$\times$}	&\textcolor{gray}{$\times$}	&\textcolor{gray}{$\times$}	&0	&0	&\textcolor{gray}{$\times$}	&\textcolor{gray}{$\times$}	&0	&0	&0	&0	&\textcolor{gray}{$\times$}	&\textcolor{gray}{$\times$}	&\textcolor{gray}{$\times$}	&\textcolor{gray}{$\times$}	&0	&0	&0	&0	\\[0.2cm]
\boldmath{$\tilde{c}_{u^2W^2D}$}	&0	&0	&0	&0	&0	&0	&0	&0	&0	&0	&0	&\textcolor{blue}{$\boldsymbol{-}$}	&\textcolor{blue}{$\boldsymbol{-}$}	&0	&0	&\textcolor{blue}{$\boldsymbol{-}$}	&\textcolor{gray}{$\times$}	&0	&0	&\textcolor{gray}{$\times$}	&\textcolor{gray}{$\times$}	&\textcolor{gray}{$\times$}	&\textcolor{gray}{$\times$}	&0	&0	&\textcolor{gray}{$\times$}	&\textcolor{gray}{$\times$}	&0	&0	&0	&0	&\textcolor{gray}{$\times$}	&\textcolor{gray}{$\times$}	&\textcolor{gray}{$\times$}	&\textcolor{gray}{$\times$}	&0	&0	&0	&0	\\[0.2cm]
\boldmath{$\tilde{c}_{u^2B^2D}$}	&0	&0	&0	&0	&0	&0	&0	&0	&0	&0	&0	&\textcolor{blue}{$\boldsymbol{-}$}	&\textcolor{blue}{$\boldsymbol{-}$}	&0	&0	&\textcolor{blue}{$\boldsymbol{-}$}	&\textcolor{gray}{$\times$}	&0	&0	&\textcolor{gray}{$\times$}	&\textcolor{gray}{$\times$}	&\textcolor{gray}{$\times$}	&\textcolor{gray}{$\times$}	&0	&0	&\textcolor{gray}{$\times$}	&\textcolor{gray}{$\times$}	&0	&0	&0	&0	&\textcolor{gray}{$\times$}	&\textcolor{gray}{$\times$}	&\textcolor{gray}{$\times$}	&\textcolor{gray}{$\times$}	&0	&0	&0	&0	\\[0.2cm]
\boldmath{$\tilde{c}_{d^2G^2D}^{(1)}$}	&0	&0	&0	&0	&0	&0	&0	&0	&0	&0	&0	&0	&0	&\textcolor{blue}{$\boldsymbol{-}$}	&\textcolor{blue}{$\boldsymbol{-}$}	&0	&0	&\textcolor{blue}{$\boldsymbol{-}$}	&\textcolor{gray}{$\times$}	&\textcolor{gray}{$\times$}	&\textcolor{gray}{$\times$}	&\textcolor{gray}{$\times$}	&\textcolor{gray}{$\times$}	&0	&0	&0	&0	&\textcolor{gray}{$\times$}	&\textcolor{gray}{$\times$}	&0	&0	&0	&0	&0	&0	&\textcolor{gray}{$\times$}	&\textcolor{gray}{$\times$}	&\textcolor{gray}{$\times$}	&\textcolor{gray}{$\times$}	\\[0.2cm]
\boldmath{$\tilde{c}_{d^2G^2D}^{(2)}$}	&0	&0	&0	&0	&0	&0	&0	&0	&0	&0	&0	&0	&0	&\textcolor{blue}{$\boldsymbol{-}$}	&\textcolor{blue}{$\boldsymbol{-}$}	&0	&0	&\textcolor{blue}{$\boldsymbol{-}$}	&\textcolor{gray}{$\times$}	&\textcolor{gray}{$\times$}	&\textcolor{gray}{$\times$}	&\textcolor{gray}{$\times$}	&\textcolor{gray}{$\times$}	&0	&0	&0	&0	&\textcolor{gray}{$\times$}	&\textcolor{gray}{$\times$}	&0	&0	&0	&0	&0	&0	&\textcolor{gray}{$\times$}	&\textcolor{gray}{$\times$}	&\textcolor{gray}{$\times$}	&\textcolor{gray}{$\times$}	\\[0.2cm]
\boldmath{$\tilde{c}_{d^2W^2D}$}	&0	&0	&0	&0	&0	&0	&0	&0	&0	&0	&0	&0	&0	&\textcolor{blue}{$\boldsymbol{-}$}	&\textcolor{blue}{$\boldsymbol{-}$}	&0	&0	&\textcolor{blue}{$\boldsymbol{-}$}	&\textcolor{gray}{$\times$}	&\textcolor{gray}{$\times$}	&\textcolor{gray}{$\times$}	&\textcolor{gray}{$\times$}	&\textcolor{gray}{$\times$}	&0	&0	&0	&0	&\textcolor{gray}{$\times$}	&\textcolor{gray}{$\times$}	&0	&0	&0	&0	&0	&0	&\textcolor{gray}{$\times$}	&\textcolor{gray}{$\times$}	&\textcolor{gray}{$\times$}	&\textcolor{gray}{$\times$}	\\[0.2cm]
\boldmath{$\tilde{c}_{d^2B^2D}$}	&0	&0	&0	&0	&0	&0	&0	&0	&0	&0	&0	&0	&0	&\textcolor{blue}{$\boldsymbol{-}$}	&\textcolor{blue}{$\boldsymbol{-}$}	&0	&0	&\textcolor{blue}{$\boldsymbol{-}$}	&\textcolor{gray}{$\times$}	&\textcolor{gray}{$\times$}	&\textcolor{gray}{$\times$}	&\textcolor{gray}{$\times$}	&\textcolor{gray}{$\times$}	&0	&0	&0	&0	&\textcolor{gray}{$\times$}	&\textcolor{gray}{$\times$}	&0	&0	&0	&0	&0	&0	&\textcolor{gray}{$\times$}	&\textcolor{gray}{$\times$}	&\textcolor{gray}{$\times$}	&\textcolor{gray}{$\times$}	\\[0.2cm]
\boldmath{$\tilde{c}_{l^2G^2D}$}	&\textcolor{blue}{$\boldsymbol{-}$}	&\textcolor{blue}{$\boldsymbol{-}$}	&0	&0	&0	&0	&\textcolor{gray}{$\times$}	&\textcolor{gray}{$\times$}	&\textcolor{gray}{$\times$}	&\textcolor{gray}{$\times$}	&0	&0	&0	&0	&0	&0	&0	&0	&0	&0	&0	&0	&0	&\textcolor{gray}{$\times$}	&\textcolor{gray}{$\times$}	&\textcolor{gray}{$\times$}	&\textcolor{gray}{$\times$}	&\textcolor{gray}{$\times$}	&\textcolor{gray}{$\times$}	&0	&0	&0	&0	&0	&0	&0	&0	&0	&0	\\[0.2cm]
\boldmath{$\tilde{c}_{l^2W^2D}$}	&\textcolor{blue}{$\boldsymbol{-}$}	&\textcolor{blue}{$\boldsymbol{-}$}	&0	&0	&0	&0	&\textcolor{gray}{$\times$}	&\textcolor{gray}{$\times$}	&\textcolor{gray}{$\times$}	&\textcolor{gray}{$\times$}	&0	&0	&0	&0	&0	&0	&0	&0	&0	&0	&0	&0	&0	&\textcolor{gray}{$\times$}	&\textcolor{gray}{$\times$}	&\textcolor{gray}{$\times$}	&\textcolor{gray}{$\times$}	&\textcolor{gray}{$\times$}	&\textcolor{gray}{$\times$}	&0	&0	&0	&0	&0	&0	&0	&0	&0	&0	\\[0.2cm]
\boldmath{$\tilde{c}_{l^2B^2D}$}	&\textcolor{blue}{$\boldsymbol{-}$}	&\textcolor{blue}{$\boldsymbol{-}$}	&0	&0	&0	&0	&\textcolor{gray}{$\times$}	&\textcolor{gray}{$\times$}	&\textcolor{gray}{$\times$}	&\textcolor{gray}{$\times$}	&0	&0	&0	&0	&0	&0	&0	&0	&0	&0	&0	&0	&0	&\textcolor{gray}{$\times$}	&\textcolor{gray}{$\times$}	&\textcolor{gray}{$\times$}	&\textcolor{gray}{$\times$}	&\textcolor{gray}{$\times$}	&\textcolor{gray}{$\times$}	&0	&0	&0	&0	&0	&0	&0	&0	&0	&0	\\[0.2cm]
\boldmath{$\tilde{c}_{e^2G^2D}$}	&0	&0	&0	&0	&0	&0	&0	&0	&0	&0	&\textcolor{blue}{$\boldsymbol{-}$}	&0	&0	&0	&0	&\textcolor{blue}{$\boldsymbol{-}$}	&\textcolor{gray}{$\times$}	&\textcolor{blue}{$\boldsymbol{-}$}	&\textcolor{gray}{$\times$}	&0	&0	&0	&0	&\textcolor{gray}{$\times$}	&\textcolor{gray}{$\times$}	&0	&0	&0	&0	&\textcolor{gray}{$\times$}	&\textcolor{gray}{$\times$}	&0	&0	&0	&0	&0	&0	&0	&0	\\[0.2cm]
\boldmath{$\tilde{c}_{e^2W^2D}$}	&0	&0	&0	&0	&0	&0	&0	&0	&0	&0	&\textcolor{blue}{$\boldsymbol{-}$}	&0	&0	&0	&0	&\textcolor{blue}{$\boldsymbol{-}$}	&\textcolor{gray}{$\times$}	&\textcolor{blue}{$\boldsymbol{-}$}	&\textcolor{gray}{$\times$}	&0	&0	&0	&0	&\textcolor{gray}{$\times$}	&\textcolor{gray}{$\times$}	&0	&0	&0	&0	&\textcolor{gray}{$\times$}	&\textcolor{gray}{$\times$}	&0	&0	&0	&0	&0	&0	&0	&0	\\[0.2cm]
\boldmath{$\tilde{c}_{e^2B^2D}$}	&0	&0	&0	&0	&0	&0	&0	&0	&0	&0	&\textcolor{blue}{$\boldsymbol{-}$}	&0	&0	&0	&0	&\textcolor{blue}{$\boldsymbol{-}$}	&\textcolor{gray}{$\times$}	&\textcolor{blue}{$\boldsymbol{-}$}	&\textcolor{gray}{$\times$}	&0	&0	&0	&0	&\textcolor{gray}{$\times$}	&\textcolor{gray}{$\times$}	&0	&0	&0	&0	&\textcolor{gray}{$\times$}	&\textcolor{gray}{$\times$}	&0	&0	&0	&0	&0	&0	&0	&0	\\[0.2cm]
\bottomrule																																								
\end{tabular}}																																								
\end{center}																																								
\caption{\it Mixing of four-fermion operators into two-fermion ones.}\label{tab:adm2}																																								
\end{table}

\begin{table}[t]																																								
\begin{center}																																								
\resizebox{1.3\textwidth}{!}{																																								
\begin{tabular}{cccccccccccccccccccccccccccccccccccccccc}																																								
\toprule																																								
 	&\boldmath{$\tilde{c}_{l^4D^2}^{(1)}$}	&\boldmath{$\tilde{c}_{l^4D^2}^{(2)}$}	&\boldmath{$\tilde{c}_{q^4D^2}^{(1)}$}	&\boldmath{$\tilde{c}_{q^4D^2}^{(2)}$}	&\boldmath{$\tilde{c}_{q^4D^2}^{(3)}$}	&\boldmath{$\tilde{c}_{q^4D^2}^{(4)}$}	&\boldmath{$\tilde{c}_{l^2q^2D^2}^{(1)}$}	&\boldmath{$\tilde{c}_{l^2q^2D^2}^{(2)}$}	&\boldmath{$\tilde{c}_{l^2q^2D^2}^{(3)}$}	&\boldmath{$\tilde{c}_{l^2q^2D^2}^{(4)}$}	&\boldmath{$\tilde{c}_{e^4D^2}$}	&\boldmath{$\tilde{c}_{u^4D^2}^{(1)}$}	&\boldmath{$\tilde{c}_{u^4D^2}^{(2)}$}	&\boldmath{$\tilde{c}_{d^4D^2}^{(1)}$}	&\boldmath{$\tilde{c}_{d^4D^2}^{(2)}$}	&\boldmath{$\tilde{c}_{e^2u^2D^2}^{(1)}$}	&\boldmath{$\tilde{c}_{e^2u^2D^2}^{(2)}$}	&\boldmath{$\tilde{c}_{e^2d^2D^2}^{(1)}$}	&\boldmath{$\tilde{c}_{e^2d^2D^2}^{(2)}$}	&\boldmath{$\tilde{c}_{u^2d^2D^2}^{(1)}$}	&\boldmath{$\tilde{c}_{u^2d^2D^2}^{(2)}$}	&\boldmath{$\tilde{c}_{u^2d^2D^2}^{(3)}$}	&\boldmath{$\tilde{c}_{u^2d^2D^2}^{(4)}$}	&\boldmath{$\tilde{c}_{l^2e^2D^2}^{(1)}$}	&\boldmath{$\tilde{c}_{l^2e^2D^2}^{(2)}$}	&\boldmath{$\tilde{c}_{l^2u^2D^2}^{(1)}$}	&\boldmath{$\tilde{c}_{l^2u^2D^2}^{(2)}$}	&\boldmath{$\tilde{c}_{l^2d^2D^2}^{(1)}$}	&\boldmath{$\tilde{c}_{l^2d^2D^2}^{(2)}$}	&\boldmath{$\tilde{c}_{q^2e^2D^2}^{(1)}$}	&\boldmath{$\tilde{c}_{q^2e^2D^2}^{(2)}$}	&\boldmath{$\tilde{c}_{q^2u^2D^2}^{(1)}$}	&\boldmath{$\tilde{c}_{q^2u^2D^2}^{(2)}$}	&\boldmath{$\tilde{c}_{q^2u^2D^2}^{(3)}$}	&\boldmath{$\tilde{c}_{q^2u^2D^2}^{(4)}$}	&\boldmath{$\tilde{c}_{q^2d^2D^2}^{(1)}$}	&\boldmath{$\tilde{c}_{q^2d^2D^2}^{(2)}$}	&\boldmath{$\tilde{c}_{q^2d^2D^2}^{(3)}$}	&\boldmath{$\tilde{c}_{q^2d^2D^2}^{(4)}$}	\\[0.2cm]
\boldmath{$\tilde{c}_{l^4D^2}^{(1)}$}	&\textcolor{gray}{$\times$}	&\textcolor{gray}{$\times$}	&0	&0	&0	&0	&\textcolor{gray}{$\times$}	&\textcolor{gray}{$\times$}	&\textcolor{gray}{$\times$}	&\textcolor{gray}{$\times$}	&0	&0	&0	&0	&0	&0	&0	&0	&0	&0	&0	&0	&0	&\textcolor{gray}{$\times$}	&\textcolor{gray}{$\times$}	&\textcolor{gray}{$\times$}	&\textcolor{gray}{$\times$}	&\textcolor{gray}{$\times$}	&\textcolor{gray}{$\times$}	&0	&0	&0	&0	&0	&0	&0	&0	&0	&0	\\[0.2cm]
\boldmath{$\tilde{c}_{l^4D^2}^{(2)}$}	&\textcolor{gray}{$\times$}	&\textcolor{gray}{$\times$}	&0	&0	&0	&0	&\textcolor{gray}{$\times$}	&\textcolor{gray}{$\times$}	&\textcolor{gray}{$\times$}	&\textcolor{gray}{$\times$}	&0	&0	&0	&0	&0	&0	&0	&0	&0	&0	&0	&0	&0	&\textcolor{gray}{$\times$}	&\textcolor{gray}{$\times$}	&\textcolor{gray}{$\times$}	&\textcolor{gray}{$\times$}	&\textcolor{gray}{$\times$}	&\textcolor{gray}{$\times$}	&0	&0	&0	&0	&0	&0	&0	&0	&0	&0	\\[0.2cm]
\boldmath{$\tilde{c}_{q^4D^2}^{(1)}$}	&0	&0	&\textcolor{gray}{$\times$}	&\textcolor{gray}{$\times$}	&\textcolor{gray}{$\times$}	&\textcolor{gray}{$\times$}	&\textcolor{gray}{$\times$}	&\textcolor{gray}{$\times$}	&\textcolor{gray}{$\times$}	&\textcolor{gray}{$\times$}	&0	&0	&0	&0	&0	&0	&0	&0	&0	&0	&0	&0	&0	&0	&0	&0	&0	&0	&0	&\textcolor{gray}{$\times$}	&\textcolor{gray}{$\times$}	&\textcolor{gray}{$\times$}	&\textcolor{gray}{$\times$}	&\textcolor{gray}{$\times$}	&\textcolor{gray}{$\times$}	&\textcolor{gray}{$\times$}	&\textcolor{gray}{$\times$}	&\textcolor{gray}{$\times$}	&\textcolor{gray}{$\times$}	\\[0.2cm]
\boldmath{$\tilde{c}_{q^4D^2}^{(2)}$}	&0	&0	&\textcolor{gray}{$\times$}	&\textcolor{gray}{$\times$}	&\textcolor{gray}{$\times$}	&\textcolor{gray}{$\times$}	&\textcolor{gray}{$\times$}	&\textcolor{gray}{$\times$}	&\textcolor{gray}{$\times$}	&\textcolor{gray}{$\times$}	&0	&0	&0	&0	&0	&0	&0	&0	&0	&0	&0	&0	&0	&0	&0	&0	&0	&0	&0	&\textcolor{gray}{$\times$}	&\textcolor{gray}{$\times$}	&\textcolor{gray}{$\times$}	&\textcolor{gray}{$\times$}	&\textcolor{gray}{$\times$}	&\textcolor{gray}{$\times$}	&\textcolor{gray}{$\times$}	&\textcolor{gray}{$\times$}	&\textcolor{gray}{$\times$}	&\textcolor{gray}{$\times$}	\\[0.2cm]
\boldmath{$\tilde{c}_{q^4D^2}^{(3)}$}	&0	&0	&\textcolor{gray}{$\times$}	&\textcolor{gray}{$\times$}	&\textcolor{gray}{$\times$}	&\textcolor{gray}{$\times$}	&\textcolor{gray}{$\times$}	&\textcolor{gray}{$\times$}	&\textcolor{gray}{$\times$}	&\textcolor{gray}{$\times$}	&0	&0	&0	&0	&0	&0	&0	&0	&0	&0	&0	&0	&0	&0	&0	&0	&0	&0	&0	&\textcolor{gray}{$\times$}	&\textcolor{gray}{$\times$}	&\textcolor{gray}{$\times$}	&\textcolor{gray}{$\times$}	&\textcolor{gray}{$\times$}	&\textcolor{gray}{$\times$}	&\textcolor{gray}{$\times$}	&\textcolor{gray}{$\times$}	&\textcolor{gray}{$\times$}	&\textcolor{gray}{$\times$}	\\[0.2cm]
\boldmath{$\tilde{c}_{q^4D^2}^{(4)}$}	&0	&0	&\textcolor{gray}{$\times$}	&\textcolor{gray}{$\times$}	&\textcolor{gray}{$\times$}	&\textcolor{gray}{$\times$}	&\textcolor{gray}{$\times$}	&\textcolor{gray}{$\times$}	&\textcolor{gray}{$\times$}	&\textcolor{gray}{$\times$}	&0	&0	&0	&0	&0	&0	&0	&0	&0	&0	&0	&0	&0	&0	&0	&0	&0	&0	&0	&\textcolor{gray}{$\times$}	&\textcolor{gray}{$\times$}	&\textcolor{gray}{$\times$}	&\textcolor{gray}{$\times$}	&\textcolor{gray}{$\times$}	&\textcolor{gray}{$\times$}	&\textcolor{gray}{$\times$}	&\textcolor{gray}{$\times$}	&\textcolor{gray}{$\times$}	&\textcolor{gray}{$\times$}	\\[0.2cm]
\boldmath{$\tilde{c}_{l^2q^2D^2}^{(1)}$}	&\textcolor{blue}{$\boldsymbol{-}$}	&\textcolor{blue}{$\boldsymbol{-}$}	&\textcolor{blue}{$\boldsymbol{-}$}	&\textcolor{blue}{$\boldsymbol{-}$}	&\textcolor{blue}{$\boldsymbol{-}$}	&\textcolor{blue}{$\boldsymbol{-}$}	&\textcolor{gray}{$\times$}	&\textcolor{gray}{$\times$}	&\textcolor{gray}{$\times$}	&\textcolor{gray}{$\times$}	&0	&0	&0	&0	&0	&0	&0	&0	&0	&0	&0	&0	&0	&\textcolor{gray}{$\times$}	&\textcolor{gray}{$\times$}	&\textcolor{gray}{$\times$}	&\textcolor{gray}{$\times$}	&\textcolor{gray}{$\times$}	&\textcolor{gray}{$\times$}	&\textcolor{gray}{$\times$}	&\textcolor{gray}{$\times$}	&\textcolor{gray}{$\times$}	&\textcolor{gray}{$\times$}	&\textcolor{gray}{$\times$}	&\textcolor{gray}{$\times$}	&\textcolor{gray}{$\times$}	&\textcolor{gray}{$\times$}	&\textcolor{gray}{$\times$}	&\textcolor{gray}{$\times$}	\\[0.2cm]
\boldmath{$\tilde{c}_{l^2q^2D^2}^{(2)}$}	&\textcolor{blue}{$\boldsymbol{-}$}	&\textcolor{blue}{$\boldsymbol{-}$}	&\textcolor{blue}{$\boldsymbol{-}$}	&\textcolor{blue}{$\boldsymbol{-}$}	&\textcolor{blue}{$\boldsymbol{-}$}	&\textcolor{blue}{$\boldsymbol{-}$}	&\textcolor{gray}{$\times$}	&\textcolor{gray}{$\times$}	&\textcolor{gray}{$\times$}	&\textcolor{gray}{$\times$}	&0	&0	&0	&0	&0	&0	&0	&0	&0	&0	&0	&0	&0	&\textcolor{gray}{$\times$}	&\textcolor{gray}{$\times$}	&\textcolor{gray}{$\times$}	&\textcolor{gray}{$\times$}	&\textcolor{gray}{$\times$}	&\textcolor{gray}{$\times$}	&\textcolor{gray}{$\times$}	&\textcolor{gray}{$\times$}	&\textcolor{gray}{$\times$}	&\textcolor{gray}{$\times$}	&\textcolor{gray}{$\times$}	&\textcolor{gray}{$\times$}	&\textcolor{gray}{$\times$}	&\textcolor{gray}{$\times$}	&\textcolor{gray}{$\times$}	&\textcolor{gray}{$\times$}	\\[0.2cm]
\boldmath{$\tilde{c}_{e^4D^2}$}	&0	&0	&0	&0	&0	&0	&0	&0	&0	&0	&\textcolor{gray}{$\times$}	&0	&0	&0	&0	&\textcolor{gray}{$\times$}	&\textcolor{gray}{$\times$}	&\textcolor{gray}{$\times$}	&\textcolor{gray}{$\times$}	&0	&0	&0	&0	&\textcolor{gray}{$\times$}	&\textcolor{gray}{$\times$}	&0	&0	&0	&0	&\textcolor{gray}{$\times$}	&\textcolor{gray}{$\times$}	&0	&0	&0	&0	&0	&0	&0	&0	\\[0.2cm]
\boldmath{$\tilde{c}_{u^4D^2}^{(1)}$}	&0	&0	&0	&0	&0	&0	&0	&0	&0	&0	&0	&\textcolor{gray}{$\times$}	&\textcolor{gray}{$\times$}	&0	&0	&\textcolor{gray}{$\times$}	&\textcolor{gray}{$\times$}	&0	&0	&\textcolor{gray}{$\times$}	&\textcolor{gray}{$\times$}	&\textcolor{gray}{$\times$}	&\textcolor{gray}{$\times$}	&0	&0	&\textcolor{gray}{$\times$}	&\textcolor{gray}{$\times$}	&0	&0	&0	&0	&\textcolor{gray}{$\times$}	&\textcolor{gray}{$\times$}	&\textcolor{gray}{$\times$}	&\textcolor{gray}{$\times$}	&0	&0	&0	&0	\\[0.2cm]
\boldmath{$\tilde{c}_{u^4D^2}^{(2)}$}	&0	&0	&0	&0	&0	&0	&0	&0	&0	&0	&0	&\textcolor{gray}{$\times$}	&\textcolor{gray}{$\times$}	&0	&0	&\textcolor{gray}{$\times$}	&\textcolor{gray}{$\times$}	&0	&0	&\textcolor{gray}{$\times$}	&\textcolor{gray}{$\times$}	&\textcolor{gray}{$\times$}	&\textcolor{gray}{$\times$}	&0	&0	&\textcolor{gray}{$\times$}	&\textcolor{gray}{$\times$}	&0	&0	&0	&0	&\textcolor{gray}{$\times$}	&\textcolor{gray}{$\times$}	&\textcolor{gray}{$\times$}	&\textcolor{gray}{$\times$}	&0	&0	&0	&0	\\[0.2cm]
\boldmath{$\tilde{c}_{d^4D^2}^{(1)}$}	&0	&0	&0	&0	&0	&0	&0	&0	&0	&0	&0	&0	&0	&\textcolor{gray}{$\times$}	&\textcolor{gray}{$\times$}	&0	&0	&\textcolor{gray}{$\times$}	&\textcolor{gray}{$\times$}	&\textcolor{gray}{$\times$}	&\textcolor{gray}{$\times$}	&\textcolor{gray}{$\times$}	&\textcolor{gray}{$\times$}	&0	&0	&0	&0	&\textcolor{gray}{$\times$}	&\textcolor{gray}{$\times$}	&0	&0	&0	&0	&0	&0	&\textcolor{gray}{$\times$}	&\textcolor{gray}{$\times$}	&\textcolor{gray}{$\times$}	&\textcolor{gray}{$\times$}	\\[0.2cm]
\boldmath{$\tilde{c}_{d^4D^2}^{(2)}$}	&0	&0	&0	&0	&0	&0	&0	&0	&0	&0	&0	&0	&0	&\textcolor{gray}{$\times$}	&\textcolor{gray}{$\times$}	&0	&0	&\textcolor{gray}{$\times$}	&\textcolor{gray}{$\times$}	&\textcolor{gray}{$\times$}	&\textcolor{gray}{$\times$}	&\textcolor{gray}{$\times$}	&\textcolor{gray}{$\times$}	&0	&0	&0	&0	&\textcolor{gray}{$\times$}	&\textcolor{gray}{$\times$}	&0	&0	&0	&0	&0	&0	&\textcolor{gray}{$\times$}	&\textcolor{gray}{$\times$}	&\textcolor{gray}{$\times$}	&\textcolor{gray}{$\times$}	\\[0.2cm]
\boldmath{$\tilde{c}_{e^2u^2D^2}^{(1)}$}	&0	&0	&0	&0	&0	&0	&0	&0	&0	&0	&\textcolor{blue}{$\boldsymbol{-}$}	&\textcolor{blue}{$\boldsymbol{-}$}	&\textcolor{blue}{$\boldsymbol{-}$}	&0	&0	&\textcolor{gray}{$\times$}	&\textcolor{gray}{$\times$}	&\textcolor{blue}{$\boldsymbol{-}$}	&\textcolor{gray}{$\times$}	&\textcolor{gray}{$\times$}	&\textcolor{gray}{$\times$}	&\textcolor{gray}{$\times$}	&\textcolor{gray}{$\times$}	&\textcolor{gray}{$\times$}	&\textcolor{gray}{$\times$}	&\textcolor{gray}{$\times$}	&\textcolor{gray}{$\times$}	&0	&0	&\textcolor{gray}{$\times$}	&\textcolor{gray}{$\times$}	&\textcolor{gray}{$\times$}	&\textcolor{gray}{$\times$}	&\textcolor{gray}{$\times$}	&\textcolor{gray}{$\times$}	&0	&0	&0	&0	\\[0.2cm]
\boldmath{$\tilde{c}_{e^2d^2D^2}^{(1)}$}	&0	&0	&0	&0	&0	&0	&0	&0	&0	&0	&\textcolor{blue}{$\boldsymbol{-}$}	&0	&0	&\textcolor{blue}{$\boldsymbol{-}$}	&\textcolor{blue}{$\boldsymbol{-}$}	&\textcolor{blue}{$\boldsymbol{-}$}	&\textcolor{gray}{$\times$}	&\textcolor{gray}{$\times$}	&\textcolor{gray}{$\times$}	&\textcolor{gray}{$\times$}	&\textcolor{gray}{$\times$}	&\textcolor{gray}{$\times$}	&\textcolor{gray}{$\times$}	&\textcolor{gray}{$\times$}	&\textcolor{gray}{$\times$}	&0	&0	&\textcolor{gray}{$\times$}	&\textcolor{gray}{$\times$}	&\textcolor{gray}{$\times$}	&\textcolor{gray}{$\times$}	&0	&0	&0	&0	&\textcolor{gray}{$\times$}	&\textcolor{gray}{$\times$}	&\textcolor{gray}{$\times$}	&\textcolor{gray}{$\times$}	\\[0.2cm]
\boldmath{$\tilde{c}_{u^2d^2D^2}^{(1)}$}	&0	&0	&0	&0	&0	&0	&0	&0	&0	&0	&0	&\textcolor{blue}{$\boldsymbol{-}$}	&\textcolor{blue}{$\boldsymbol{-}$}	&\textcolor{blue}{$\boldsymbol{-}$}	&\textcolor{blue}{$\boldsymbol{-}$}	&\textcolor{blue}{$\boldsymbol{-}$}	&\textcolor{gray}{$\times$}	&\textcolor{blue}{$\boldsymbol{-}$}	&\textcolor{gray}{$\times$}	&\textcolor{gray}{$\times$}	&\textcolor{gray}{$\times$}	&\textcolor{gray}{$\times$}	&\textcolor{gray}{$\times$}	&0	&0	&\textcolor{gray}{$\times$}	&\textcolor{gray}{$\times$}	&\textcolor{gray}{$\times$}	&\textcolor{gray}{$\times$}	&0	&0	&\textcolor{gray}{$\times$}	&\textcolor{gray}{$\times$}	&\textcolor{gray}{$\times$}	&\textcolor{gray}{$\times$}	&\textcolor{gray}{$\times$}	&\textcolor{gray}{$\times$}	&\textcolor{gray}{$\times$}	&\textcolor{gray}{$\times$}	\\[0.2cm]
\boldmath{$\tilde{c}_{u^2d^2D^2}^{(2)}$}	&0	&0	&0	&0	&0	&0	&0	&0	&0	&0	&0	&\textcolor{blue}{$\boldsymbol{-}$}	&\textcolor{blue}{$\boldsymbol{-}$}	&\textcolor{blue}{$\boldsymbol{-}$}	&\textcolor{blue}{$\boldsymbol{-}$}	&\textcolor{blue}{$\boldsymbol{-}$}	&\textcolor{gray}{$\times$}	&\textcolor{blue}{$\boldsymbol{-}$}	&\textcolor{gray}{$\times$}	&\textcolor{gray}{$\times$}	&\textcolor{gray}{$\times$}	&\textcolor{gray}{$\times$}	&\textcolor{gray}{$\times$}	&0	&0	&\textcolor{gray}{$\times$}	&\textcolor{gray}{$\times$}	&\textcolor{gray}{$\times$}	&\textcolor{gray}{$\times$}	&0	&0	&\textcolor{gray}{$\times$}	&\textcolor{gray}{$\times$}	&\textcolor{gray}{$\times$}	&\textcolor{gray}{$\times$}	&\textcolor{gray}{$\times$}	&\textcolor{gray}{$\times$}	&\textcolor{gray}{$\times$}	&\textcolor{gray}{$\times$}	\\[0.2cm]
\boldmath{$\tilde{c}_{l^2e^2D^2}^{(1)}$}	&\textcolor{blue}{$\boldsymbol{-}$}	&\textcolor{blue}{$\boldsymbol{-}$}	&0	&0	&0	&0	&\textcolor{gray}{$\times$}	&\textcolor{gray}{$\times$}	&\textcolor{gray}{$\times$}	&\textcolor{gray}{$\times$}	&\textcolor{blue}{$\boldsymbol{-}$}	&0	&0	&0	&0	&\textcolor{blue}{$\boldsymbol{-}$}	&\textcolor{gray}{$\times$}	&\textcolor{blue}{$\boldsymbol{-}$}	&\textcolor{gray}{$\times$}	&0	&0	&0	&0	&\textcolor{gray}{$\times$}	&\textcolor{gray}{$\times$}	&\textcolor{gray}{$\times$}	&\textcolor{gray}{$\times$}	&\textcolor{gray}{$\times$}	&\textcolor{gray}{$\times$}	&\textcolor{gray}{$\times$}	&\textcolor{gray}{$\times$}	&0	&0	&0	&0	&0	&0	&0	&0	\\[0.2cm]
\boldmath{$\tilde{c}_{l^2u^2D^2}^{(1)}$}	&\textcolor{blue}{$\boldsymbol{-}$}	&\textcolor{blue}{$\boldsymbol{-}$}	&0	&0	&0	&0	&\textcolor{gray}{$\times$}	&\textcolor{gray}{$\times$}	&\textcolor{gray}{$\times$}	&\textcolor{gray}{$\times$}	&0	&\textcolor{blue}{$\boldsymbol{-}$}	&\textcolor{blue}{$\boldsymbol{-}$}	&0	&0	&\textcolor{blue}{$\boldsymbol{-}$}	&\textcolor{gray}{$\times$}	&0	&0	&\textcolor{gray}{$\times$}	&\textcolor{gray}{$\times$}	&\textcolor{gray}{$\times$}	&\textcolor{gray}{$\times$}	&\textcolor{gray}{$\times$}	&\textcolor{gray}{$\times$}	&\textcolor{gray}{$\times$}	&\textcolor{gray}{$\times$}	&\textcolor{gray}{$\times$}	&\textcolor{gray}{$\times$}	&0	&0	&\textcolor{gray}{$\times$}	&\textcolor{gray}{$\times$}	&\textcolor{gray}{$\times$}	&\textcolor{gray}{$\times$}	&0	&0	&0	&0	\\[0.2cm]
\boldmath{$\tilde{c}_{l^2d^2D^2}^{(1)}$}	&\textcolor{blue}{$\boldsymbol{-}$}	&\textcolor{blue}{$\boldsymbol{-}$}	&0	&0	&0	&0	&\textcolor{gray}{$\times$}	&\textcolor{gray}{$\times$}	&\textcolor{gray}{$\times$}	&\textcolor{gray}{$\times$}	&0	&0	&0	&\textcolor{blue}{$\boldsymbol{-}$}	&\textcolor{blue}{$\boldsymbol{-}$}	&0	&0	&\textcolor{blue}{$\boldsymbol{-}$}	&\textcolor{gray}{$\times$}	&\textcolor{gray}{$\times$}	&\textcolor{gray}{$\times$}	&\textcolor{gray}{$\times$}	&\textcolor{gray}{$\times$}	&\textcolor{gray}{$\times$}	&\textcolor{gray}{$\times$}	&\textcolor{gray}{$\times$}	&\textcolor{gray}{$\times$}	&\textcolor{gray}{$\times$}	&\textcolor{gray}{$\times$}	&0	&0	&0	&0	&0	&0	&\textcolor{gray}{$\times$}	&\textcolor{gray}{$\times$}	&\textcolor{gray}{$\times$}	&\textcolor{gray}{$\times$}	\\[0.2cm]
\boldmath{$\tilde{c}_{q^2e^2D^2}^{(1)}$}	&0	&0	&\textcolor{blue}{$\boldsymbol{-}$}	&\textcolor{blue}{$\boldsymbol{-}$}	&\textcolor{blue}{$\boldsymbol{-}$}	&\textcolor{blue}{$\boldsymbol{-}$}	&\textcolor{gray}{$\times$}	&\textcolor{gray}{$\times$}	&\textcolor{gray}{$\times$}	&\textcolor{gray}{$\times$}	&\textcolor{blue}{$\boldsymbol{-}$}	&0	&0	&0	&0	&\textcolor{blue}{$\boldsymbol{-}$}	&\textcolor{gray}{$\times$}	&\textcolor{blue}{$\boldsymbol{-}$}	&\textcolor{gray}{$\times$}	&0	&0	&0	&0	&\textcolor{gray}{$\times$}	&\textcolor{gray}{$\times$}	&0	&0	&0	&0	&\textcolor{gray}{$\times$}	&\textcolor{gray}{$\times$}	&\textcolor{gray}{$\times$}	&\textcolor{gray}{$\times$}	&\textcolor{gray}{$\times$}	&\textcolor{gray}{$\times$}	&\textcolor{gray}{$\times$}	&\textcolor{gray}{$\times$}	&\textcolor{gray}{$\times$}	&\textcolor{gray}{$\times$}	\\[0.2cm]
\boldmath{$\tilde{c}_{q^2u^2D^2}^{(1)}$}	&0	&0	&\textcolor{blue}{$\boldsymbol{-}$}	&\textcolor{blue}{$\boldsymbol{-}$}	&\textcolor{blue}{$\boldsymbol{-}$}	&\textcolor{blue}{$\boldsymbol{-}$}	&\textcolor{gray}{$\times$}	&\textcolor{gray}{$\times$}	&\textcolor{gray}{$\times$}	&\textcolor{gray}{$\times$}	&0	&\textcolor{blue}{$\boldsymbol{-}$}	&\textcolor{blue}{$\boldsymbol{-}$}	&0	&0	&\textcolor{blue}{$\boldsymbol{-}$}	&\textcolor{gray}{$\times$}	&0	&0	&\textcolor{gray}{$\times$}	&\textcolor{gray}{$\times$}	&\textcolor{gray}{$\times$}	&\textcolor{gray}{$\times$}	&0	&0	&\textcolor{gray}{$\times$}	&\textcolor{gray}{$\times$}	&0	&0	&\textcolor{gray}{$\times$}	&\textcolor{gray}{$\times$}	&\textcolor{gray}{$\times$}	&\textcolor{gray}{$\times$}	&\textcolor{gray}{$\times$}	&\textcolor{gray}{$\times$}	&\textcolor{gray}{$\times$}	&\textcolor{gray}{$\times$}	&\textcolor{gray}{$\times$}	&\textcolor{gray}{$\times$}	\\[0.2cm]
\boldmath{$\tilde{c}_{q^2u^2D^2}^{(2)}$}	&0	&0	&\textcolor{blue}{$\boldsymbol{-}$}	&\textcolor{blue}{$\boldsymbol{-}$}	&\textcolor{blue}{$\boldsymbol{-}$}	&\textcolor{blue}{$\boldsymbol{-}$}	&\textcolor{gray}{$\times$}	&\textcolor{gray}{$\times$}	&\textcolor{gray}{$\times$}	&\textcolor{gray}{$\times$}	&0	&\textcolor{blue}{$\boldsymbol{-}$}	&\textcolor{blue}{$\boldsymbol{-}$}	&0	&0	&\textcolor{blue}{$\boldsymbol{-}$}	&\textcolor{gray}{$\times$}	&0	&0	&\textcolor{gray}{$\times$}	&\textcolor{gray}{$\times$}	&\textcolor{gray}{$\times$}	&\textcolor{gray}{$\times$}	&0	&0	&\textcolor{gray}{$\times$}	&\textcolor{gray}{$\times$}	&0	&0	&\textcolor{gray}{$\times$}	&\textcolor{gray}{$\times$}	&\textcolor{gray}{$\times$}	&\textcolor{gray}{$\times$}	&\textcolor{gray}{$\times$}	&\textcolor{gray}{$\times$}	&\textcolor{gray}{$\times$}	&\textcolor{gray}{$\times$}	&\textcolor{gray}{$\times$}	&\textcolor{gray}{$\times$}	\\[0.2cm]
\boldmath{$\tilde{c}_{q^2d^2D^2}^{(1)}$}	&0	&0	&\textcolor{blue}{$\boldsymbol{-}$}	&\textcolor{blue}{$\boldsymbol{-}$}	&\textcolor{blue}{$\boldsymbol{-}$}	&\textcolor{blue}{$\boldsymbol{-}$}	&\textcolor{gray}{$\times$}	&\textcolor{gray}{$\times$}	&\textcolor{gray}{$\times$}	&\textcolor{gray}{$\times$}	&0	&0	&0	&\textcolor{blue}{$\boldsymbol{-}$}	&\textcolor{blue}{$\boldsymbol{-}$}	&0	&0	&\textcolor{blue}{$\boldsymbol{-}$}	&\textcolor{gray}{$\times$}	&\textcolor{gray}{$\times$}	&\textcolor{gray}{$\times$}	&\textcolor{gray}{$\times$}	&\textcolor{gray}{$\times$}	&0	&0	&0	&0	&\textcolor{gray}{$\times$}	&\textcolor{gray}{$\times$}	&\textcolor{gray}{$\times$}	&\textcolor{gray}{$\times$}	&\textcolor{gray}{$\times$}	&\textcolor{gray}{$\times$}	&\textcolor{gray}{$\times$}	&\textcolor{gray}{$\times$}	&\textcolor{gray}{$\times$}	&\textcolor{gray}{$\times$}	&\textcolor{gray}{$\times$}	&\textcolor{gray}{$\times$}	\\[0.2cm]
\boldmath{$\tilde{c}_{q^2d^2D^2}^{(2)}$}	&0	&0	&\textcolor{blue}{$\boldsymbol{-}$}	&\textcolor{blue}{$\boldsymbol{-}$}	&\textcolor{blue}{$\boldsymbol{-}$}	&\textcolor{blue}{$\boldsymbol{-}$}	&\textcolor{gray}{$\times$}	&\textcolor{gray}{$\times$}	&\textcolor{gray}{$\times$}	&\textcolor{gray}{$\times$}	&0	&0	&0	&\textcolor{blue}{$\boldsymbol{-}$}	&\textcolor{blue}{$\boldsymbol{-}$}	&0	&0	&\textcolor{blue}{$\boldsymbol{-}$}	&\textcolor{gray}{$\times$}	&\textcolor{gray}{$\times$}	&\textcolor{gray}{$\times$}	&\textcolor{gray}{$\times$}	&\textcolor{gray}{$\times$}	&0	&0	&0	&0	&\textcolor{gray}{$\times$}	&\textcolor{gray}{$\times$}	&\textcolor{gray}{$\times$}	&\textcolor{gray}{$\times$}	&\textcolor{gray}{$\times$}	&\textcolor{gray}{$\times$}	&\textcolor{gray}{$\times$}	&\textcolor{gray}{$\times$}	&\textcolor{gray}{$\times$}	&\textcolor{gray}{$\times$}	&\textcolor{gray}{$\times$}	&\textcolor{gray}{$\times$}	\\[0.2cm]
\bottomrule																																								
\end{tabular}}																																								
\end{center}																																								
\caption{\it Mixing between four-fermion operators.}\label{tab:adm4}																																								
\end{table}

\end{landscape}


\section{Some cross-checks}
\label{app:xchecks}
To further cross-check the validity of our approach towards restricting the ADM of the dimension-eight SMEFT, we have computed explicitly a number of elements of this matrix based on the approach of Ref.~\cite{Li:2023edf}. Some exmaples are:
\begin{align}
  4\gamma_{\tilde{c}_{q^2\phi^2D^3}^{(1)}\,,\,\tilde{c}_{q^4 D^2}^{(1)}}& =  \gamma_{\tilde{c}_{q^2\phi^2D^3}^{(1)}\,,\,\tilde{c}_{q^4 D^2}^{(3)}} = 2\gamma_{\tilde{c}_{q^2\phi^2D^3}^{(2)}\,,\,\tilde{c}_{q^4 D^2}^{(2)}} = \gamma_{\tilde{c}_{q^2\phi^2D^3}^{(2)}\,,\,\tilde{c}_{q^4 D^2}^{(4)}} = -16 |Y_u|^2\,, \\
  2\gamma_{\tilde{c}_{q^2\phi^2D^3}^{(1)}\,,\,\tilde{c}_{q^4 D^2}^{(2)}}& =  \gamma_{\tilde{c}_{q^2\phi^2D^3}^{(1)}\,,\,\tilde{c}_{q^4 D^2}^{(4)}} = 4\gamma_{\tilde{c}_{q^2\phi^2D^3}^{(2)}\,,\,\tilde{c}_{q^4 D^2}^{(1)}} = \gamma_{\tilde{c}_{q^2\phi^2D^3}^{(2)}\,,\,\tilde{c}_{q^4 D^2}^{(3)}} = -16|Y_d|^2\,,\\
 4\gamma_{\tilde{c}_{q^2B^2D^1}^{(1)}\,,\,\tilde{c}_{q^4 D^2}^{(1)}}& = 2\gamma_{\tilde{c}_{q^2B^2D^1}^{(1)}\,,\,\tilde{c}_{q^4 D^2}^{(2)}} = \gamma_{\tilde{c}_{q^2B^2D^1}^{(1)}\,,\,\tilde{c}_{q^4 D^2}^{(3)}}=\gamma_{\tilde{c}_{q^2B^2D^1}^{(1)}\,,\,\tilde{c}_{q^4 D^2}^{(4)}}=-\frac{8}{27}g_1^2\,,\\
 \gamma_{\tilde{c}_{l^2\phi^2D^3}^{(1)}\,,\,\tilde{c}_{l^4 D^2}^{(1)}}& = 2\gamma_{\tilde{c}_{l^2\phi^2D^3}^{(2)}\,,\,\tilde{c}_{l^4 D^2}^{(2)}} = -4 |Y_l|^2\,, \\
 \gamma_{\tilde{c}_{l^2B^2D^1}^{(1)}\,,\,\tilde{c}_{l^4 D^2}^{(1)}}& = 2\gamma_{\tilde{c}_{l^2B^2D^1}^{(1)}\,,\,\tilde{c}_{l^4 D^2}^{(2)}} = -\frac{4}{3}g_1^2\,,\\
 4\gamma_{\tilde{c}_{u^2\phi^2D^3}^{(1)}\,,\,\tilde{c}_{u^4D^2}^{(1)}} &= \gamma_{\tilde{c}_{u^2\phi^2D^3}^{(1)}\,,\,\tilde{c}_{u^4D^2}^{(2)}} = -16|Y_u|^2\,,\\
 4\gamma_{\tilde{c}_{u^2B^2D}\,,\,\tilde{c}_{u^4D^2}^{(1)}} &= \gamma_{\tilde{c}_{u^2B^2D}\,,\,\tilde{c}_{u^4D^2}^{(2)}} = -\frac{128}{27}g_1^2\,,\\
 \gamma_{\tilde{c}_{l^2B^2D}^{(1)}\,,\,\tilde{c}_{l^2\phi^2D^3}^{(1)}} &= \gamma_{\tilde{c}_{l^2B^2D}^{(1)}\,,\,\tilde{c}_{l^2\phi^2D^3}^{(2)}} = -\frac{1}{24}g_1^2\,,\\
 \gamma_{\tilde{c}_{l^2B^2D}^{(1)}\,,\,\tilde{c}_{l^2\phi^2D^3}^{(3)}} &= \gamma_{\tilde{c}_{l^2B^2D}^{(1)}\,,\,\tilde{c}_{l^2\phi^2D^3}^{(4)}} = \fcolorbox{black}{LightCyan}{\,0\,}\,,\\
 \gamma_{\tilde{c}_{e^2B^2D}\,,\,\tilde{c}_{e^2\phi^2D^3}^{(1)}} &= -\frac{1}{3}g_1^2\,,\\
 \gamma_{\tilde{c}_{e^2B^2D}\,,\,\tilde{c}_{e^2\phi^2D^3}^{(2)}} &= \fcolorbox{black}{LightCyan}{\,0\,}\,.
\end{align}
They all agree with the results in Tabs.~\ref{tab:adm1}--\ref{tab:adm4}.


\bibliographystyle{style} 

\bibliography{refs} 

\end{document}